\documentclass[prc,amsmath,amssymb,superscriptaddress,floatfix,nofootinbib,11pt]{revtex4}

\usepackage{inputenc}
\usepackage{amstext,amsfonts}
\usepackage{mathrsfs}
\usepackage{bbm,bm}
\usepackage{slashed}
\usepackage{etoolbox}
\robustify{\slashed}
\usepackage{setspace}
\usepackage{graphicx}
\usepackage{subfig}
\usepackage{appendix}
\usepackage{float}
\usepackage{array}
\usepackage{braket}
\usepackage[colorlinks=true, allcolors=blue]{hyperref}
\usepackage[capitalize]{cleveref}

\DeclareUnicodeCharacter{2212}{-}

\usepackage{xcolor}

\makeatletter
\def\@dotsep{4.5}
\makeatother

\newcommand{\simle}{\hspace*{0.2em}\raisebox{0.5ex}{$<$}
     \hspace{-0.8em}\raisebox{-0.3em}{$\sim$}\hspace*{0.2em}}

\begin{document}

\title{Two-Pion Exchange Contributions to the Nucleon-Nucleon Interaction \\
from the  Roper Resonance}
\author{Yang Xiao}
\affiliation{School of Space and Environment, Beihang University, Beijing 102206, China}
\affiliation{School of Physics, Beihang University, Beijing 102206, China}
\affiliation{Center for Exotic Nuclear Studies, Institute for Basic Science, Daejeon 34126, Republic of Korea}

\author{Li-Sheng~Geng}
\email[]{Email: lisheng.geng@buaa.edu.cn}
\affiliation{School of Physics, Beihang University, Beijing 102206, China}
 \affiliation{Sino-French Carbon Neutrality Research Center, \'Ecole Centrale de P\'ekin/School of General Engineering, Beihang University, Beijing 100191, China}
\affiliation{Peng Huanwu Collaborative Center for Research and Education, Beihang University, Beijing 100191, China}
\affiliation{Southern Center for Nuclear-Science Theory (SCNT), Institute of Modern Physics, Chinese Academy of Sciences, Huizhou 516000, China}

\author{U. van Kolck}
\email[]{Email: vankolck@ijclab.in2p3.fr}
\affiliation{European Centre for Theoretical Studies in Nuclear Physics and Related Areas (ECT*), Fondazione Bruno Kessler, 38123 Villazzano (TN), Italy}
\affiliation{Université Paris-Saclay, CNRS/IN2P3, IJCLab, 91405 Orsay, France}
\affiliation{Department of Physics, University of Arizona, Tucson, AZ 85721, USA}

\begin{abstract}
We derive the long-range components of the nucleon-nucleon (NN) two-pion-exchange potential with an intermediate Roper resonance. Leading-order interactions in heavy-baryon chiral perturbation theory are considered. To assess the Roper contributions in a unified framework, we include the leading contributions from the Delta resonance, as well as mixed Delta-Roper effects. NN phase shifts with orbital angular momentum $L\geq 2$ are calculated in first-order perturbation theory and compared to those obtained without the  Roper resonance.  We find that the Delta sets the dominant scale of the resonance-induced corrections, while the Roper-dependent effects are smaller, strongly channel dependent, and most visible in the $D$ waves.
\end{abstract}

\maketitle

\section{Introduction}

Nuclear interactions are a fundamental input in nuclear physics and astrophysics, which have been comprehensively studied in effective field theories (EFTs) of QCD since the early
1990s~\cite{Weinberg:1990rz,Weinberg:1991um,Ordonez:1992xp,Weinberg:1992yk,Ordonez:1993tn,vanKolck:1994yi,Ordonez:1995rz} --- see Ref.~\cite{Hammer:2019poc} for a recent review. In Chiral EFT ($\chi$EFT), nuclear forces are of two types. The lightness of the pion, guaranteed by spontaneous chiral-symmetry breaking and small quark masses, makes it responsible for long-range interactions through one or more exchanges. The short-range interactions encode details of QCD dynamics at distances $r\simle M_{\rm QCD}^{-1}$, where $M_{\rm QCD}\sim 1$ GeV is the characteristic nonperturbative scale of the theory associated with most baryon masses, such as the nucleon's $m_N\simeq 940$ MeV. While determining the low-energy constants (LECs) associated with contact interactions requires nuclear input, the long-range potential can be calculated using LECs obtained from one-baryon processes. Here, we examine a novel long-range component of the nucleon-nucleon (NN) interaction.

Long-range NN forces are expected to be amenable to the power counting employed in $\chi$EFT restricted to at most one nucleon, chiral perturbation theory~\cite{Weinberg:1978kz}. The leading-order (LO) long-range potential comes from the well-known static one-pion exchange (OPE), with two-pion exchange (TPE) as the first correction. Since the first derivations~\cite{Ordonez:1993tn,Ordonez:1995rz,Kaiser:1997mw,Kaiser:1998wa}, much work has been devoted to calculations of these forces in an explicit nonrelativistic expansion based on heavy-baryon chiral perturbation theory (HB$\chi$PT)~\cite{Jenkins:1990jv,Jenkins:1991es}. Together with a prescription for the ordering of short-range interactions, OPE and TPE have led to many phenomenologically successful chiral potentials as reviewed in, for example, Refs.~\cite{Machleidt:2011zz,Epelbaum:2019kcf,Hammer:2019poc}. In recent years, a relativistically covariant EFT has been proposed and found to converge faster than these conventional chiral
potentials~\cite{Ren:2016jna,Xiao:2018jot,Xiao:2020ozd,Bai:2020yml,Wang:2020myr,Ren:2017yvw,Wang:2021kos,Bai:2021uim,Lu:2021gsb,Xiao:2024jmu,Lu:2025syk,Xiao:2025ufy,Lu:2025ubc}.

Most existing chiral nuclear forces contain only the ground-state nucleon, with the contributions from baryon excited states relegated to LECs at higher orders. State-of-the-art chiral nuclear forces have been constructed up to the fifth order in the nonrelativistic formalism~\cite{Entem:2017gor} and up to the third order in the covariant formalism~\cite{Lu:2021gsb}. Although already of high precision, these chiral nuclear forces still encounter some problems, {\it e.g.}, convergence with respect to the components of the TPE potential. It has long been realized~\cite{Ordonez:1993tn,Ordonez:1995rz,Kaiser:1997mw} that the formally subleading component of TPE is comparable or even larger than the presumably leading TPE, which challenges the power counting underlying $\chi$EFT. It has even been proposed~\cite{Valderrama:2010aw,PavonValderrama:2019lsu,Mishra:2021luw,Valderrama:2021bql} that TPE should be promoted to LO together, or even replace, OPE.

The convergence issues of chiral nuclear forces can be ameliorated by explicitly including baryon resonances. If the effects of a resonance are relegated to LECs, non-analytic contributions are lost, which limits the expansion range of the theory. Including a low-lying resonance explicitly as a low-energy degree of freedom not only extends the validity of the EFT but also reorganizes the multipion-exchange contributions and improves order-by-order convergence. The lowest nucleon excitation is the Delta isobar, $\Delta$(1232), whose mass $m_\Delta$ is only $\delta\equiv m_\Delta-m_N\simeq 300$ MeV higher than the nucleon's.
Its contribution to
NN~\cite{Ordonez:1993tn,Ordonez:1995rz,Kaiser:1998wa,Krebs:2007rh,Epelbaum:2008td,Piarulli:2014bda,Ekstrom:2017koy,Strohmeier:2020dkb,Nosyk:2021pxb} and three-nucleon~\cite{vanKolck:1994yi,Pandharipande:2005sx,Epelbaum:2007sq,Kaiser:2015yca,Krebs:2018jkc} chiral forces has attracted increasing attention. Delta effects, which would otherwise appear only in the subleading TPE via LECs, become manifest already in the leading TPE. Moreover, some of the nuclear matter properties are also found to be better described with Deltaful chiral potentials~\cite{Piarulli:2014bda,Ekstrom:2017koy,Logoteta:2016nzc,Nosyk:2021pxb}.

In TPE, a nucleon scatters a virtual pion. $\chi$EFT can be extended to the Delta region in pion-nucleon scattering~\cite{Long:2009wq} following a similar approach to other hadronic reactions~\cite{Pascalutsa:2002pi}. However, the $P_{11}$ channel still converges poorly, unless the second nucleon excitation --- the Roper resonance $N(1440)$~\cite{Roper:1964zza}, lying $\rho\equiv m_R-m_N\simeq 500$ MeV above the nucleon --- is introduced explicitly~\cite{Long:2011rt}. Unlike higher resonances, the Roper width is compatible with having an origin in the chiral expansion. The Roper also nearly saturates the Adler-Weisberger sum rule together with nucleon and Delta, which suggests that these baryons fall into a simple reducible representation of the chiral group~\cite{Weinberg:1969hw,Beane:2002ud}.
By now, several properties of the Roper have been studied in $\chi$EFT~\cite{Borasoy:2006fk,Djukanovic:2009gt,Bauer:2014cqa,Gegelia:2016xcw,Gelenava:2017mmk,Severt:2020jzc}.

The role of the Roper in chiral nuclear forces is explored here for the first time~\footnote{After the first version of our paper appeared, a complementary spectral-function analysis was presented~\cite{Kaiser:2026lvl} of two-pion-exchange NN potentials with the Roper excitation, including single- and double-Roper intermediate states as well as combined Delta-Roper excitations. The corresponding momentum-space potentials were constructed through subtracted dispersion relations with spectral-function regularization.}. Previous calculations of Roper effects in the NN system predate $\chi$EFT. They were based on coupled-channel approaches, with an emphasis on the pion-production region and above; see, for example, Refs.~\cite{Lomon:1981su,Lee:1984us,Ray:1987ir,Sitarski:1987gd}.
Here, we derive in HB$\chi$PT the TPE NN potentials with one and two intermediate Ropers, together with the pure Delta and mixed Delta-Roper contributions, using dimensional regularization. This potential is to be employed in conjunction with other components of the chiral nuclear force, with an eye to future extensions into the pion-production region.
Our main focus here remains the Roper, but the inclusion of the much more studied $\Delta(1232)$ isobar and of mixed Delta-Roper diagrams allows us to assess the relative importance of Roper effects in a unified framework.
As we are going to see, the pure Roper contributions are relatively small at low energies, while the Delta and mixed Delta-Roper contributions provide useful benchmarks for their size.
We compare the Roper, Delta, and mixed Delta-Roper TPE potentials with the leading TPE containing only nucleons in intermediate states
and calculate their effects on the NN elastic scattering phase shifts for partial waves with orbital angular momentum $L \geq 2$,
where the centrifugal barrier favors the use of perturbation theory.
The contribution from intermediate Roper resonances is found to be strongly channel dependent and sizeable for $D$ waves.

The manuscript is organized as follows. In Sec.~\ref{SecDerivation}, we present the effective Lagrangians involving the nucleon, the $\Delta(1232)$ isobar, and the $N(1440)$, and derive the corresponding leading TPE NN potentials in dimensional regularization. The isoscalar and isovector TPE potentials in momentum space and the resulting NN scattering phase shifts for $L\geq 2$ are shown and discussed in Sec.~\ref{SecResults}. A summary and outlook are given in Sec.~\ref{SecOutlook}.

\section{Roper contributions to nucleon-nucleon scattering}
\label{SecDerivation}

In this section, we derive the long-range components of the leading TPE potential in the presence of intermediate Delta isobar and Roper resonances. After the relevant terms in the Lagrangian are presented in Sec.~\ref{SubSecLag}, the pure-Delta, mixed Delta-Roper and pure-Roper triangle and box topologies are examined in Sec.~\ref{SubSecDiag}.

\subsection{Effective Lagrangian}
\label{SubSecLag}

The Roper has the same quantum numbers (spin $S=1/2$ and isospin $I=1/2$) as the nucleon and therefore both can be represented by isospinor fields, which we denote as $R$ and $N$, respectively. Since we are interested in the nonrelativistic regime where antibaryons only contribute to short-range effects, we can use heavy particle fields~\cite{Jenkins:1990jv,Jenkins:1991es}, each with two components representing particle spin states. The Delta resonance, in contrast, has $S=3/2$ and $I=3/2$ and is represented by a field $T_\mu^i$ that incorporates four spin states for each isospin component. The inert nucleon mass is removed through a phase redefinition, leaving behind the Roper-nucleon and Delta-nucleon mass differences, $\rho\equiv m_R-m_N$ and $\delta\equiv m_\Delta-m_N$, respectively. The baryon covariant velocity is denoted by $v_\mu$ and the covariant spin operator is $S_\mu = i \gamma_5 \sigma_{\mu\nu} v^{\nu}/2$. The projection on the physical Delta degrees of freedom is enforced by
\begin{equation}
P_{3/2}^{\mu\nu}=g^{\mu\nu}-v^\mu v^\nu-\frac{4}{1-d}S^\mu S^\nu,
\label{eq:delta-building-blocks0}
\end{equation}
where $d$ is the spacetime dimension. We will work in the baryons' rest frame, where $v_\mu = (1,0,0,0)$ and $S_\mu=(0, \vec{\sigma}/2)$ in terms of the Pauli spin matrices $\vec{\sigma}$. The contributions from antibaryons are represented in the EFT by contact interactions.

In addition, we focus on momenta comparable to the pion mass, where the isospin-triplet $\boldsymbol{\pi}$ of pseudoscalar pions needs to be considered explicitly. Pions are pseudo-Goldstone bosons of spontaneous chiral SU(2)$\times$ SU(2) symmetry breaking, characterized by the pion decay constant $f_\pi\simeq 92$ MeV. Chiral symmetry is realized nonlinearly~\cite{Weinberg:1968de}, for example through the use of a field~\cite{Coleman:1969sm,Callan:1969sn}
$U(x) = u^2(x) = \exp{(i \boldsymbol{\pi}\cdot \boldsymbol{\tau}/f_\pi)}$, where $\boldsymbol{\tau}$ are the Pauli matrices in isospin space.
Chiral symmetry is implemented easily through chiral-covariant derivatives,
\begin{align}
u_\mu &=
i\left(
u^\dagger\partial_\mu u
-u\partial_\mu u^\dagger
\right),
\\
w_\mu^i&=\frac{1}{2}\,\mathrm{Tr}(\tau^i u_\mu),
\\
\Gamma_\mu &=
\frac{1}{2}
\left(
u^\dagger\partial_\mu u
+u\partial_\mu u^\dagger
\right),
\\
\Gamma_\mu^i
&\equiv
\frac{1}{2}\operatorname{Tr}\!\left(\tau^i\Gamma_\mu\right),
\\
D_\mu\Psi &=
\left(\partial_\mu+\Gamma_\mu\right)\Psi,
\qquad \Psi=N,R,
\\
D_\mu^{ij}T_\nu^j &=
\left[
\delta^{ij}
\left(\partial_\mu+\Gamma_\mu\right)
-2i\epsilon^{ijk}\Gamma_\mu^k
\right]T_\nu^j,
\end{align}
for the pion field, a spin-1/2 field $\Psi$, and the Delta field.

To calculate the leading TPE potential with the Delta isobar and
Roper resonance in HB$\chi$PT, we need the leading terms in the effective chiral Lagrangian including pions, nucleon, Delta, and Roper~\cite{Long:2009wq, Long:2011rt},
\begin{align}
\mathcal{L}^{(0)}={}&
\frac{f_\pi^2}{4}\,\mathrm{Tr}\!\left[(\partial_\mu U)(\partial^\mu U^\dag)
+m_\pi^2(U+U^\dag)\right]
+\bar N\left(iv\!\cdot\!D+g_A S\!\cdot\!u\right)N
\nonumber\\
&+\bar R\left(iv\!\cdot\!D-\rho\right)R
-\bar T_\mu^i\left(iv\!\cdot\!D^{ij}-\delta ~\delta^{ij}\right)g^{\mu\nu}T_\nu^j
\nonumber\\
&+g_A'\left(\bar R S\!\cdot\!u\,N+\text{H.c.}\right)
+\frac{h_A}{2}\left(\bar T_\mu^i P_{3/2}^{\mu\nu}w_\nu^i N+\text{H.c.}\right).
\label{Lag}
\end{align}
The $\pi NN$, $\pi N\Delta$, and $\pi NR$ couplings are denoted by $g_A$, $h_A$, and $g_A'$, respectively. We use the normalization in which the leading $\pi N\Delta$ vertex is proportional to $h_A/2f_\pi$. Subleading terms include more derivatives, powers of the pion mass, and powers of the mass splittings. Among the higher-order terms, one also finds isospin-breaking terms proportional to the quark-mass difference and the electromagnetic fine-structure constant. With naive dimensional analysis~\cite{Manohar:1983md}, which incorporates naturalness in a perturbative context~\cite{vanKolck:2020plz}, higher-order terms are suppressed by powers of $M_{\rm QCD}\sim \Lambda_\chi \equiv 4\pi f_\pi$.

\subsection{Two-pion exchange potential with intermediate Deltas and Ropers}
\label{SubSecDiag}

The nucleon propagator from the Lagrangian~\eqref{Lag} is static. Two-nucleon reducible diagrams --- that is, those that can be split by cutting only intermediate nucleon lines --- have a pinch singularity and are infrared divergent, which requires~\cite{Weinberg:1991um} a resummation in the nucleon propagator of the recoil term proportional to $m_N^{-1}$ appearing in the Lagrangian with an additional derivative, ${\cal L}^{(1)}$. These diagrams are then infrared finite but enhanced by a factor of~\cite{Bedaque:2002mn,Hammer:2019poc} $4\pi m_N/p$, where $p$ is the characteristic momentum of the process. The NN potential is defined as the sum of two-nucleon irreducible diagrams. Being free of the infrared divergence, contributions to the long-range potential can be estimated~\cite{Weinberg:1991um} with the same power counting of $\chi$PT~\cite{Weinberg:1978kz}. Neglecting isospin-breaking corrections, the long-range potential is a perturbative expansion in $Q/M_{\rm QCD}$, where $Q$ represents the low-energy scales: $p$, $m_\pi$, $\delta$, and in our case  $\rho$ as well.

\begin{figure}[tb]
\includegraphics[width=0.9\textwidth]{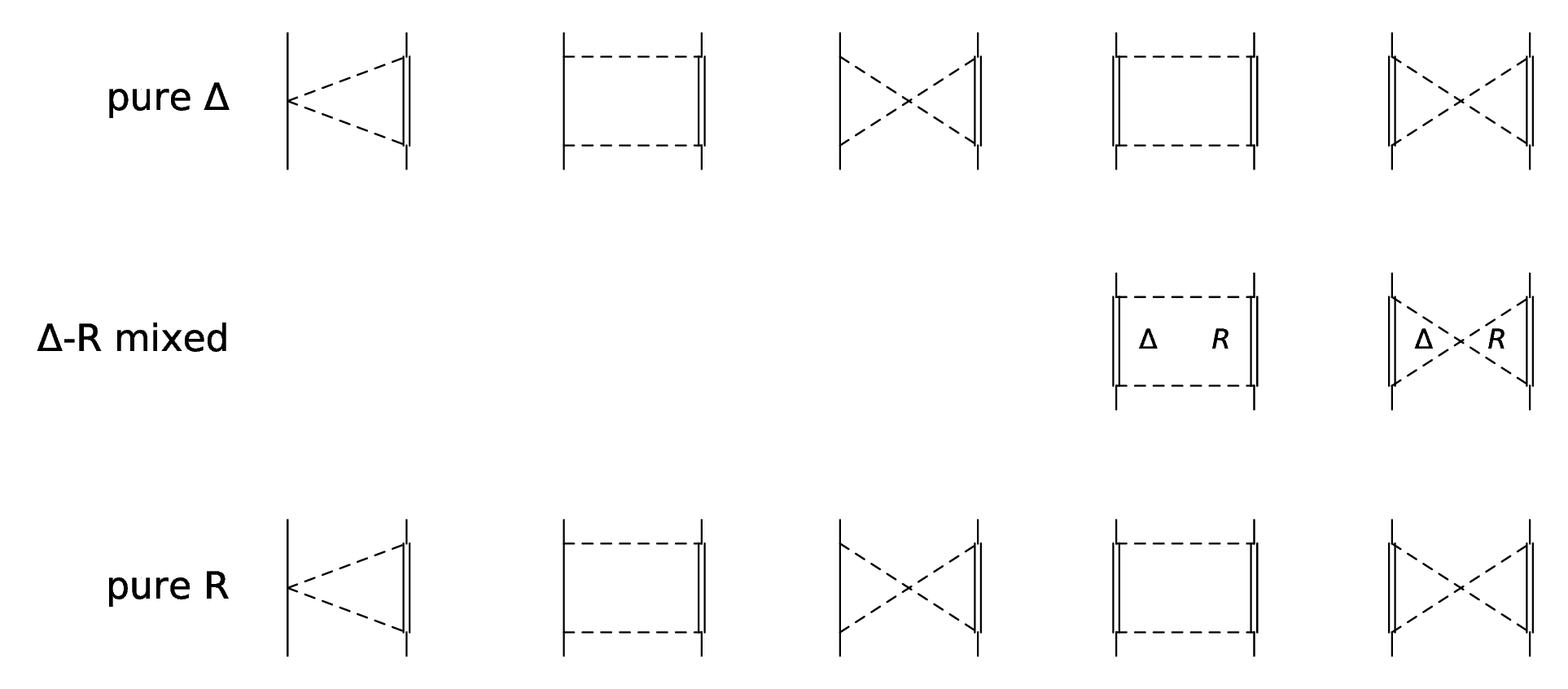}
\caption{\label{fig:all}Leading two-pion-exchange topologies with pure Delta, mixed Delta-Roper, and pure Roper intermediate states, shown from top to bottom. Solid and dashed lines denote nucleons and pions, respectively, while double lines denote baryon resonances. In the mixed diagrams, $\Delta$ and $R$ label the Delta and Roper propagators. The corresponding mirror diagrams are not shown. Propagators and vertices are from the Lagrangian $\mathcal{L}^{(0)}$, Eq.~\eqref{Lag}.}
\end{figure}

The leading contributions to the long-range NN potential from intermediate Delta isobars and Roper resonances  arise from the Feynman diagrams shown in~\cref{fig:all}. These are all the one-loop diagrams with resonances one can build with the propagators and vertices extracted from the leading Lagrangian in Eq.~\eqref{Lag}. They represent contributions in Roperful $\chi$EFT of ${\cal O}(Q^2/M_{\rm QCD}^2)$ relative to OPE. Since there are no pinch singularities, baryons are static with recoil corrections expected at the next order. The Delta and Roper decay widths stemming from loop corrections to the corresponding propagators enter two orders down in the chiral expansion. The triangle diagrams arise from the Weinberg-Tomozawa vertex contained in the nucleon chiral-covariant derivative and are proportional to $h_A^2$ or $g_A'^2$. The planar- and crossed-box diagrams are proportional to $g_A^2 h_A^2$, $g_A^2 g_A'^2$, $h_A^4$, $h_A^2g_A'^2$, or $g_A'^4$, depending on whether they contain one or two intermediate Deltas/Ropers. Note that there is always at least one resonance in intermediate states, and no subtraction of iterated OPE is needed. Thus, in the limit $\rho\to 0$ and $g_A'\to g_A$ the mixed Delta-Roper diagrams reduce to one-Delta planar boxes but the pure-Roper TPE potential differs from the nucleon-only potential at the same order. At next order, relative ${\cal O}(Q/M_{\rm QCD})$, there are further contributions to the TPE potential where the Weinberg-Tomozawa vertex is replaced by pion-nucleon interactions not fixed by chiral symmetry, whose LECs in Deltaless $\chi$EFT are often denoted $c_i$. Treating the Delta and Roper explicitly, as we do here, accounts for the largest part of these contributions.

As with any other loop diagrams, those in Fig.~\ref{fig:all} require regularization. They contain short-range components that do not vanish when the regulator is removed. This regulator dependence can be eliminated by appropriately regulating the bare strengths of two-nucleon contact interactions. The asymptotic form of the potential is obtained at distances much larger than the inverse of the momentum cutoff; see, for example, Refs.~\cite{Ordonez:1995rz,Friar:1999sj} for nucleon and Delta intermediate states. However, the same result is most cleanly obtained with dimensional regularization and minimal subtraction, which we employ in the following. The calculation is analogous to that in Refs.~\cite{Kaiser:1997mw,Kaiser:1998wa}.

We consider the diagrams in Fig.~\ref{fig:all} in the center-of-mass (CM) frame and denote the initial and final momenta by $\vec{p}$ and $\vec{p}\,'$, respectively, with $\vec{q}=|\vec{q}\,|\, \hat q = \vec{p}-\vec{p} \,'$ the transferred momentum.
We associate with it the static four-momentum transfer $q^\mu=(0,\vec q)$, with $q^2\equiv q_\mu q^\mu=-\vec q^{\,2}$.
Representing the spin and isospin operators of nucleon $i$ by $\vec{\sigma}_i$ and $\boldsymbol{\tau}_i$, the corresponding potential takes the form
\begin{align}
    V = \ V_C +  W_C \, \boldsymbol{\tau}_1 \cdot \boldsymbol{\tau}_2
    + \left(V_T + W_T \, \boldsymbol{\tau}_1 \cdot \boldsymbol{\tau}_2 \right) \,
            \left(\vec{\sigma}_1 \cdot \hat{q} \ \vec{\sigma}_2 \cdot \hat{q}-\vec{\sigma}_1\cdot\vec{\sigma}_2
    \right),
    \label{Vstructure}
\end{align}
where the subscripts $C$  and $T$ refer to the central  and quasi-tensor components, respectively, with $V$ denoting the isoscalar potential and $W$ the isovector potential.
Note that the quasi-tensor component includes not only the tensor $S_{12}(\hat{q}) \equiv \vec{\sigma}_1 \cdot \hat{q} \ \vec{\sigma}_2 \cdot \hat{q}-\vec{\sigma}_1\cdot \vec{\sigma}_2/3$ but also a spin-spin component.

The calculation is straightforward but long, involving combinations of several integrals with dimension-dependent coefficients. After subtraction of the poles in $d-4$, one is left with functions that are nonanalytic in the momentum as well as polynomial terms that depend on the renormalization scale $\mu$. The quasi-tensor components arise exclusively from box diagrams in several combinations involving nucleon-Delta, nucleon-Roper, and Delta-Roper integrals. The central components emerge not only from boxes but also from triangle diagrams involving only a Delta or a Roper resonance. We can write the sum of the one- and two-Delta boxes to the Delta triangle as
\begin{align}
V_C^{(\Delta)}={}&
\frac{g_A^2h_A^2}{4f_\pi^4}
\left(\mathcal C_B^{\Delta N}(q)+\mathcal C_X^{\Delta N}(q)\right)
+\frac{h_A^4}{12f_\pi^4}
\left(\mathcal D_B^{\Delta\Delta}(q)
+\mathcal D_X^{\Delta\Delta}(q)\right),
\nonumber\\
W_C^{(\Delta)}={}&
-\frac{h_A^2}{6f_\pi^4}\,\mathcal S^\Delta(q)
+\frac{g_A^2h_A^2}{12f_\pi^4}
\left(\mathcal C_B^{\Delta N}(q)-\mathcal C_X^{\Delta N}(q)\right)
+\frac{h_A^4}{72f_\pi^4}
\left(-\mathcal D_B^{\Delta\Delta}(q)
+\mathcal D_X^{\Delta\Delta}(q)\right),
\nonumber\\
V_T^{(\Delta)}={}&
-\frac{g_A^2h_A^2}{4f_\pi^4}\, q^2
\mathcal K_-^{\Delta N}(q)
+\frac{h_A^4}{12f_\pi^4}\, q^2
\mathcal K_-^{\Delta\Delta}(q),
\nonumber\\
W_T^{(\Delta)}={}&
-\frac{g_A^2h_A^2}{12f_\pi^4}\,q^2
\mathcal K_+^{\Delta N}(q)
-\frac{h_A^4}{72f_\pi^4}\, q^2
\mathcal K_+^{\Delta\Delta}(q),
\label{eq:delta-total-main}
\end{align}
the complete pure-Roper contribution as
\begin{align}
V_C^{(R)}={}&
\frac{3g_A^2g_A'^2}{8f_\pi^4}
\left(\mathcal R_B^{RN}(q)+\mathcal R_X^{RN}(q)\right)
+\frac{3g_A'^4}{16f_\pi^4}
\left(\mathcal R_B^{RR}(q)+\mathcal R_X^{RR}(q)\right),
\nonumber\\
W_C^{(R)}={}&
\frac{g_A'^2}{2f_\pi^4}\,\mathcal T^R(q)
+\frac{g_A^2g_A'^2}{4f_\pi^4}
\left(-\mathcal R_B^{RN}(q)+\mathcal R_X^{RN}(q)\right)
+\frac{g_A'^4}{8f_\pi^4}
\left(-\mathcal R_B^{RR}(q)+\mathcal R_X^{RR}(q)
\right),
\nonumber\\
V_T^{(R)}={}&
\frac{3g_A^2g_A'^2}{8f_\pi^4}\, q^2
\mathcal K_-^{RN}(q)
+\frac{3g_A'^4}{16f_\pi^4}\, q^2
\mathcal K_-^{RR}(q),
\nonumber\\
W_T^{(R)}={}&
-\frac{g_A^2g_A'^2}{4f_\pi^4}\, q^2
\mathcal K_+^{RN}(q)
-\frac{g_A'^4}{8f_\pi^4}\, q^2
\mathcal K_+^{RR}(q),
\label{eq:roper-total-main}
\end{align}
and the mixed Delta-Roper boxes as
\begin{align}
V_C^{(\Delta R)}={}&
\frac{g_A'^2h_A^2}{4f_\pi^4}
\left(\mathcal C_B^{\Delta R}(q)+\mathcal C_X^{\Delta R}(q)
\right),
\nonumber\\
W_C^{(\Delta R)}={}&
\frac{g_A'^2h_A^2}{12f_\pi^4}
\left(\mathcal C_B^{\Delta R}(q)-\mathcal C_X^{\Delta R}(q)\right),
\nonumber\\
V_T^{(\Delta R)}={}&
-\frac{g_A'^2h_A^2}{4f_\pi^4}\, q^2
\mathcal K_-^{\Delta R}(q),
\nonumber\\
W_T^{(\Delta R)}={}&
-\frac{g_A'^2h_A^2}{12f_\pi^4}\, q^2
\mathcal K_+^{\Delta R}(q).
\label{eq:mixed-total-main}
\end{align}
The functions $\mathcal S^{\Delta}$ and $\mathcal T^R$ for triangle diagrams and $\mathcal K^{\Delta N, \Delta\Delta, RN, RR, \Delta R}_{\pm}$, $\mathcal C^{\Delta N,\Delta R}_{B,X}$, $\mathcal R^{RN,RR}_{B,X}$, and $\mathcal D^{\Delta\Delta}_{B,X}$ for planar ($B$) and crossed ($X$) boxes are given in Appendix~\ref{app:loop-functions}.

This potential is local because it depends on momenta only through the momentum transfer $\vec{q}$. It contains analytic terms that cannot be separated from LECs. The necessity of LECs is clear from the explicit dependence of the momentum functions on the renormalization scale $\mu$, shown in Appendix~\ref{app:loop-functions}.
Had we used a momentum cutoff $\Lambda$, it would appear instead of $\mu$ and make it evident that the separation is arbitrary between TPE and short-range interactions at the same order. There are two (seven) independent contact interactions we can construct with no (two) powers of momenta~\cite{Ordonez:1992xp,Ordonez:1995rz}. Accounting for explicit chiral-symmetry breaking, two more contact interactions appear with two powers of the pion mass. These LECs can remove all the analytic terms
in Eqs.~\eqref{eq:delta-total-main}$-$\eqref{eq:mixed-total-main}.
The $\mu$ dependence in quasi-tensor terms can be removed by the LECs of two two-derivative, while for central terms the LECs are needed from two other two-derivative contact interactions, the two chiral-breaking interactions, and the two non-derivative interactions present already at leading order.

More interesting are the nonanalytic terms in $|\vec{q}|$, which cannot be reproduced by a finite series of short-range interactions. They are independent of the regulator: they can be obtained instead for a momentum cutoff when $\Lambda\to \infty$. These non-analytic contributions display the expected suppression of $Q^2/M_{\rm QCD}^2$ compared to OPE in the form of dimensional coefficients $\propto$ $\delta^2$, $\rho^2, m_\pi^2$, $\vec{q}^{\,2}$ divided by $(4\pi f_\pi)^2$.
As a check on the analytical expressions, we analytically continued the dimensionally regularized amplitudes from spacelike momentum transfer,  $q^2=-\vec q^{\,2}$, to the timelike region, $q^2=\mu^2+i0$, or equivalently $|\vec q|=i\mu-0^+$, with $\mu>2m_\pi$, and compared the resulting imaginary parts, or spectral functions, with Refs.~\cite{Kaiser:1998wa,Kaiser:2026lvl}. After accounting for the overall sign and tensor-operator conventions, we find agreement topology by topology and component by component for the pure-Delta, pure-Roper, and mixed Delta--Roper sectors. This comparison verifies the nonanalytic, finite-range parts of the potentials. The remaining additive analytic polynomial terms are scheme-dependent and can be absorbed into short-range contact interactions.

The form of the potential and its effects on observables are studied in the next section.

\section{Numerical Results and Discussion}
\label{SecResults}

We first specify the numerical input and the subtraction used to display the momentum-space potentials. We then discuss the cumulative effects of the nucleon, Delta, and Roper TPE contributions to peripheral NN phase shifts.

\subsection{Values of low-energy constants}
\label{SecLECs}

To calculate the NN scattering phase shifts, we must fix all relevant LECs. The LO nucleon axial-vector coupling is fixed at $g_A =1.29$, the pion decay constant at $f_\pi=92.4$ MeV, the pion mass at $m_\pi=139$ MeV, and the nucleon mass at $m_N=939$ MeV. The exact values of these LECs are not very important, as the Roper parameters are much less well known and their uncertainties dominate the uncertainty of our results.

For illustration, we use values for the Delta and Roper sectors obtained from fits to $\pi N$ scattering in Ref.~\cite{Long:2011rt}, which extends $\chi$EFT to the Delta region, includes an explicit Roper field, and uses the nucleon parameters above. Specifically, we use $h_A=2.92$ for the nucleon-Delta coupling and $\delta=305$ MeV for the Delta-nucleon mass splitting. These values correspond to the same $\pi N\Delta$ coupling convention as in Eq.~\eqref{Lag}.

For the nucleon-Roper coupling $g_A'$ and mass splitting $\rho$ we adopt the values listed in Table I of Ref.~\cite{Long:2011rt}. Note that $\rho$ is the residual mass parameter entering the Roper propagator in HB$\chi$PT rather than the Breit-Wigner mass. We display the two pairs of values from the HB$\chi$PT analysis: the ``N$^2$LO GW'' values $(g_A',\rho)=(0.54,550~\text{MeV})$ and the ``N$^2$LO EFT'' values $(g_A',\rho)=(1.06,690~\text{MeV})$. The ``N$^2$LO EFT" values are obtained from a fit to the GW phase-shift analysis (PSA)~\cite{Arndt:2006bf} near the Delta resonance. In this region, Delta effects are enhanced by two orders, so in the Roper-nucleon sector ``N$^2$LO EFT'' includes only the parameters we consider here. This set of LECs is therefore consistent from an EFT perspective. However, these values are quite far from those corresponding to a reducible representation of the chiral group~\cite{Weinberg:1969hw,Beane:2002ud}, which is suggested by the saturation of the Adler-Weisberger sum rule. To provide a measure of the sensitivity to the Roper input, we include also the ``N$^2$LO GW'' values, which are fitted at the same order considered here --- that is, ${\cal O}(Q^2/M_{\rm QCD}^2)$ with respect to OPE ---
but to the values of the Breit-Wigner mass and width extracted from the GW PSA~\cite{Arndt:2006bf}. For this set of values, the Roper effects are enhanced by the smaller mass splitting but reduced by the smaller couplings. The labels ``Roper (GW)'' and ``Roper (EFT)'' used in the figures indicate only which of these two pairs of values is employed. For each curve, the same pair is used in both the pure-Roper and mixed Delta-Roper contributions. All other ingredients of the calculation are unchanged. The separation of the two curves is therefore only a sensitivity test of the Roper input, not a statistical confidence interval or an EFT truncation band.

\subsection{The potential as a function of momentum transfer}

The potential itself is not observable, as its short-range part is regulator dependent to ensure that the full amplitude, from which observables are extracted, is properly renormalized. The separation between analytic short-range terms in the TPE potential and contact interactions depends on the renormalization convention. Nevertheless, the long-range part of the potential (given by the non-analytic terms) provides some intuition about its effects on the full amplitude. Here we compare the long-range TPE effects of the resonances with the TPE with nucleons only, which is formally leading in Deltaless $\chi$EFT and calculated in the same framework~\cite{Kaiser:1997mw}.

To expose the finite-range three-momentum dependence, we subtract polynomials in $|\vec q|^2$ from the result of the ``raw'' potential components obtained above by a loop calculation. Since the potential is two orders beyond leading order where non-derivative contact interactions first appear, subtract a polynomial linear in $|\vec q|^2$ from each raw
component $Z=V_{C,T},W_{C,T}$:
\begin{align}
Z^{\rm sub}(|\vec q|)
=
Z^{\rm raw}(|\vec q|)
-\left(a_0^{Z}+a_1^{Z}|\vec q|^2\right),
\label{eq:poly-subtraction}
\end{align}
where
\begin{align}
a_0^Z&=\lim_{|\vec q|^2\to0}Z^{\rm raw}(|\vec q|),
&
a_1^Z&=\lim_{|\vec q|^2\to0}\frac{\partial Z^{\rm raw}(|\vec q|)}{\partial |\vec q|^2}.
\label{eq:poly-coefficients}
\end{align}
For the quasi-tensor components, which start linearly in $|\vec{q}|^2$, $a_0^{V_T}=a_0^{W_T}=0$. For the central components the constant coefficients do not necessarily vanish since they can be absorbed in the LECs of non-derivative contact interactions, which appear already at leading order.

After subtraction, the renormalization scale $\mu$ disappears. As a check, we have found no noticeable dependence when we numerically varied $\mu$ from 0.5 to 2.0 GeV in steps of 0.5 GeV for the Roper ``N$^2$LO EFT" case. The subtraction coefficients themselves are $\mu$ dependent. They are listed in Appendix~\ref{app:polynomial-subtractions} for $\mu=1$ GeV as an illustration. 

\begin{figure}[tb]
\includegraphics[width=1.0\textwidth]{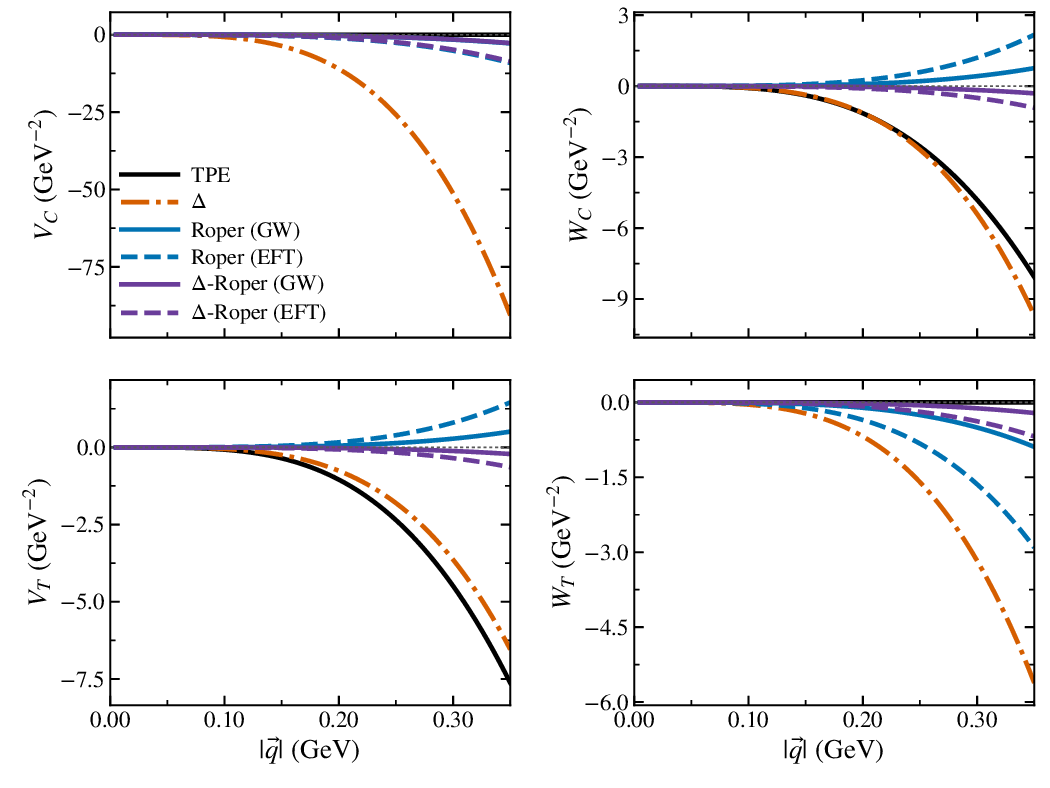}
\caption{Subtracted isoscalar (left column) and isovector (right column) components of the TPE NN potential, in GeV$^{-2}$, as functions of momentum transfer $|\vec q|$ in GeV. The upper, and lower rows show the central components, and quasi-tensor components, respectively. The black solid curves denote the leading pure-nucleon TPE, the orange dash-dotted curves pure-Delta TPE, the blue solid and dashed curves pure-Roper TPE, and the purple solid and dashed curves mixed Delta-Roper TPE. Solid and dashed resonance curves use the ``N$^2$LO GW" and ``N$^2$LO EFT" Roper parameter pairs, respectively. The curves are separate contributions, not cumulative sums. The subtraction is defined in Eqs.~\eqref{eq:poly-subtraction} and~\eqref{eq:poly-coefficients},
and the numerical coefficients are given in Appendix~\ref{app:polynomial-subtractions}.}
\label{fig:isopotential}
\end{figure}

In~\cref{fig:isopotential} we compare, as a function of the momentum transfer  $|\vec{q}|$, the various components of the TPE potential after subtraction: pure-nucleon, pure-Delta, pure-Roper, and mixed Delta-Roper, the last two with the two sets of parameters discussed above.

Figure~\ref{fig:isopotential} exhibits a clear hierarchy.
For all components, the pure-Delta contribution is large, being attractive in the central and quasi-tensor channels. It is comparable to, and reinforces, pure-nucleon TPE in the channels (isovector central and isoscalar tensor) where the latter does not vanish. The Delta also generates a large and attractive isoscalar central potential. These large contributions support the inclusion of Delta as an explicit degree of freedom in $\chi$EFT.

In contrast, the Roper contributions are smaller, the Delta controlling the overall resonance scale in most channels. The figure also shows why the Roper effects cannot be assessed consistently without the Delta: the mixed term can either reinforce or cancel the pure-Roper term depending on the spin-isospin component. They reinforce each other in $V_C$ and $W_T$ but partially cancel in $W_C$ and $V_T$. The difference between the ``GW" and ``EFT" curves is most pronounced in the Roper contributions to the isovector components and in the mixed contribution to $V_C$. The ``N$^2$LO EFT" parameters give the largest contributions. The largest effects are in the isoscalar central and isovector quasi-tensor potentials. They are sizable for $W_T$, where the pure-Roper contribution is roughly half of the pure-Delta's and reinforces it, the effect being enhanced by the mixed contribution.

\subsection{Comparison with Roperless potentials}

When the Roper is integrated out, its propagator is replaced by analytic terms suppressed by powers of $\{|\vec{q}|,m_\pi\}/\rho$ in addition to $\{\vec{q}^{\,2},m_\pi^2\}/(4\pi f_\pi)^2$. The corresponding contributions will enter TPE at higher orders, with the largest effects expected from the smallest number of intermediate Ropers. The pure one-Roper contributions proportional to $1/\rho$ reduce to
\begin{align}
      V_C^{(1 \slashed{R})} = &- \frac{3 g_A^2 g_A'^2 }{64 \pi  f_\pi^4  \rho}\left[m_\pi  \omega^2(|\vec{q}|)+\left(2 m_\pi^2+ \vec{q}^{\,2}\right)^2 A(|\vec q|) \right],
      \label{eq:totpotVC_sat}\\
      W_C^{(1 \slashed{R})} = & \; 0,
      \label{eq:totpotWC_sat}\\
      V_T^{(1 \slashed{R})}
            =& \; 0,
      \label{eq:totpotVT_sat}\\
      W_T^{(1 \slashed{R})}
            =& -\frac{g_A^2 g_A'^2 \vec{q}^{\,2}}{64 \pi  f_\pi^4 \rho } \left[ m_\pi +\omega^2(|\vec q|) A(|\vec q|)\right],
      \label{eq:totpotWT_sat}
\end{align}
where
\begin{equation}
      \omega(|\vec q|)=\sqrt{4m_\pi^2+\vec q^{\,2}}, \qquad
A(|\vec q|)=\frac{1}{2|\vec q|}\tan^{-1}\!\frac{|\vec q|}{2m_\pi}.
\end{equation}
Actually, the above equations for $V_C$ and $W_T$ are exact. As expected, this potential has the form of the subleading TPE in Deltaless $\chi$EFT
with the substitution of $g_A'^2/2 \rho \rightarrow c_4 = -2c_3$. The relation between $c_3$ and $c_4$ is the same as that shown in Ref.~\cite{Bernard:1996gq} for the Roper saturation, wherein $c_4^R = -2c_3^R= g_A^2 \tilde{R}/8 \rho$ with $\tilde{R}=0.28$. Notice a value of $g_A' \approx 0.34 $ $(\rho = 470$ MeV) is suggested by the Roper saturation by comparing the two relations of $c_3$ and $c_4$. Such a value is consistent with the ``N$^3$LO EFT" value listed in Table I of Ref.~\cite{Long:2011rt}. The ``N$^2$LO EFT" values $g_A' =1.06$ and $\rho=690$ MeV suggest instead that $c_3^R \approx -0.40 $ GeV$^{-1}$ and $c_4^R \approx 0.81$ GeV$^{-1}$. Their values are much larger than those shown in Ref.~\cite{Bernard:1996gq} ($c_3^R=-0.06$ GeV$^{-1}$ and $c_4^R=0.12$ GeV$^{-1}$), implying an enhancement in the contribution to the potential from the intermediate Roper resonance at this order.

The subleading two-pion-nucleon interactions include not only the momentum-dependent interactions with LECs $c_3$ and $c_4$, but also energy- and pion-mass-dependent interactions with LECs $c_2$ and $c_1$, respectively. In the conventions used here, the Roper-saturated LECs satisfy
\begin{equation}
c_1^R=0,
\qquad
c_2^R=-c_3^R=\frac{c_4^R}{2}
=\frac{g_A'^2}{4\rho}.
\label{eq:roper-saturated-ci}
\end{equation}
The $c_2^R$ interaction does not contribute to the static Deltaless TPE potential at this order. It must, however, be retained when a Delta remains explicit, because the nonzero Delta--nucleon mass splitting makes the corresponding pion-energy dependence contribute to the nonpolynomial part of the potential.

The mixed Delta--Roper diagrams can therefore be matched onto a Delta triangle containing the Roper-saturated $\pi\pi NN$ vertex. The leading integrated-out mixed Delta-Roper potential is
\begin{align}
V_C^{(\Delta\slashed R)}(|\vec q|)
\doteq{}&
-\frac{g_A'^2h_A^2\delta}
{288\pi^2f_\pi^4\rho}
\biggl\{
6\Sigma_\Delta^2(|\vec q|)
D_\Delta(|\vec q|)
+\left(
12\delta^2-20m_\pi^2-11\vec q^{\,2}
\right)L_\Delta(|\vec q|)
\biggr\},
\nonumber\\
W_C^{(\Delta\slashed R)}(|\vec q|)
={}&0,
\nonumber\\
V_T^{(\Delta\slashed R)}(|\vec q|)
={}&0,
\nonumber\\
W_T^{(\Delta\slashed R)}(|\vec q|)
\doteq{}&
-\frac{g_A'^2h_A^2\delta\,\vec q^{\,2}}
{576\pi^2f_\pi^4\rho}
\biggl\{
\left[
\omega^2(|\vec q|)-4\delta^2
\right]D_\Delta(|\vec q|)
-2L_\Delta(|\vec q|)
\biggr\}.
\label{eq:delta-roper-integrated-out}
\end{align}
Here $\doteq$ denotes equality up to a polynomial $a_0+a_1\vec q^{\,2}$, which is removed by the subtraction defined in Eqs.~\eqref{eq:poly-subtraction} and~\eqref{eq:poly-coefficients}, and
\begin{align}
L_\Delta(|\vec q|)
&\equiv
\frac{\omega(|\vec q|)}{|\vec q|}
\ln\left[
\frac{\omega(|\vec q|)+|\vec q|}{2m_\pi}
\right],
\nonumber\\
\Sigma_\Delta(|\vec q|)
&\equiv
2m_\pi^2+\vec q^{\,2}-2\delta^2,
\nonumber\\
D_\Delta(|\vec q|)
&\equiv
\frac{1}{\delta}
\int_{2m_\pi}^{\infty}
\frac{d\mu}{\mu^2+\vec q^{\,2}}
\tan^{-1}
\frac{\sqrt{\mu^2-4m_\pi^2}}{2\delta}.
\label{eq:delta-loop-functions-integrated-out}
\end{align}
This potential is obtained from the general Delta-triangle expressions proportional to the LECs $c_i$ given in Eq.~(2.9) of Ref.~\cite{Krebs:2007rh} with Eq.~\eqref{eq:roper-saturated-ci}.
We have checked that expanding the explicit-Roper expression in inverse powers of $\rho$ and retaining only the leading term agrees with this result within numerical precision.

Together with the Deltaless contributions in Eqs.~\eqref{eq:totpotVC_sat}--\eqref{eq:totpotWT_sat}, Eq.~\eqref{eq:delta-roper-integrated-out} gives the leading one-Roper contribution in the Deltaful theory with the Roper integrated out. At this order, only the $V_C$ and $W_T$ components are nonvanishing, explaining the larger magnitudes of these components when the Roper is kept explicit. To assess the accuracy of this description, we define
\begin{equation}
\mathcal R_Z(|\vec q|)
\equiv
\frac{
Z^{(1\slashed R) {\rm sub}}(|\vec q|)
+
Z^{(\Delta\slashed R) {\rm sub}}(|\vec q|)
}{
Z^{(R) {\rm sub}}(|\vec q|)
+
Z^{(\Delta R) {\rm sub}}(|\vec q|)
},
\qquad
Z=V_C,W_T .
\label{eq:roper-integrated-out-ratio}
\end{equation}
The numerator contains the integrated-out nucleon- and Delta-triangle contributions, whereas the denominator contains the complete
potential with explicit intermediate Ropers. Here the superscript ``sub'' indicates that each contribution entering Eq.~\eqref{eq:roper-integrated-out-ratio} is subjected to the same polynomial subtraction defined in Eqs.~\eqref{eq:poly-subtraction} and~\eqref{eq:poly-coefficients} before the ratio is formed. Since the subtraction is linear, it can be applied either to the individual contributions or to their respective sums. The numerical subtraction coefficients for the combined numerator and denominator are collected in Appendix~\ref{app:polynomial-subtractions} for $\mu=1$ GeV as an illustration.

\begin{figure}[htbp]
  \centering
  \includegraphics[width=\columnwidth]
    {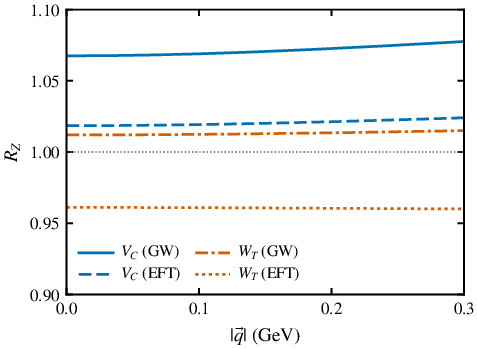}
  \caption{Ratios $\mathcal R_Z$ between integrated-out- and explicit-Roper potentials, defined in Eq.~\eqref{eq:roper-integrated-out-ratio}, as functions of the momentum transfer $|\vec q|$.  The curves correspond to the ``N$^2$LO GW'' and ``N$^2$LO EFT'' Roper parameter sets.}
  \label{fig:roper-integrated-out-ratio}
\end{figure}

As shown in Fig.~\ref{fig:roper-integrated-out-ratio}, the ratios depend only weakly on the momentum transfer and remain within approximately $10\%$ of unity for both Roper parameter sets. Thus, over the momentum range considered here, the leading long-range Roper effects can be encoded in the LECs of the Deltaful theory with only modest changes.

\subsection{Phase shifts and mixing angles of high partial waves}

To quantify the effects of the Roper on observables, we calculate NN scattering in high partial waves. The short-range part of our potential, which cannot be separated from contact interactions in a model-independent way, contributes only to waves with orbital angular momentum $L\le 1$. To focus on the long-range part of the TPE potential, which can be attributed to the Delta and Roper, we consider phase shifts with $L\ge 2$ and mixing angles with $J\ge 2$.

The NN scattering phase shifts and mixing angles can be obtained by using the ``Stapp'' --- or ``bar'' --- parameterization~\cite{Stapp:1956mz} of the $S$ matrix, which is related to the on-shell $T$ matrix. To obtain the on-shell $T$ matrix, one must insert the potential into a scattering equation, such as the Lippmann-Schwinger equation. However, for large angular momenta, the centrifugal barrier effectively reduces the effects of the potential by inverse factors of $L$. The suppression is stronger for shorter-range components of the potential, as the centrifugal barrier more effectively shields them. As a consequence, partial waves with $L\geq 2$ probe predominantly the long-range components of the potential, which can be treated perturbatively~\cite{Nogga:2005hy,Birse:2005um,PavonValderrama:2016lqn,Wu:2018lai,Kaplan:2019znu,Peng:2020nyz}.
For these waves, phase shifts are essentially monotonic and relatively small, $|\delta_L| \simle 30^\circ$, in which case $\tan\delta_L\approx \delta_L$. The difference between phase shifts obtained in the scattering equation and in perturbation theory is then small. The Delta and Roper effects can be seen in the phase shifts and mixing angles calculated in first-order perturbation theory,
\begin{align}
    \delta_{LSJ} = &
    -\frac{m_N^2 p}{4\pi E}\; \text{Re} \bra {LSJ} V \ket{LSJ},
    \\
    \varepsilon_J = &-
    \frac{m_N^2 p}{4\pi E}\; \text{Re} \bra {J-1, 1,J} V \ket{J+1,1,J}.
\end{align}
where $p=|\vec{p}|$ and $E=\sqrt{m_N^2+p^2}$ are the CM
momentum and energy, respectively, and $\ket{LSJ}$ is the free wavefunction in the basis of well-defined orbital angular momentum, spin, and total angular momentum. On-shell, the transferred momentum $|\vec{q}\,|=p\sqrt{2(1-z)}$, $z=\cos\theta$, with $\theta$ the scattering angle. The exclusion principle constrains the isospin $I$ so that
$I+S+L$ is odd. The matrix elements can be written as angular integrals of the isospin-projected combinations
\begin{align}
   U_{C,T} = V_{C,T}+ (4I-3) W_{C,T},
\end{align}
multiplied by Legendre polynomials of degree $J$ or $J\pm 1$, $P_J(z)$ or $P_{J\pm 1}(z)$.
The explicit partial-wave projections can be found in Ref.~\cite{Kaiser:1997mw}~\footnote{To use those formulas with Eq.~\eqref{Vstructure}, one must translate the convention of Ref.~\cite{Kaiser:1997mw}: if its spin-spin and unnormalized tensor coefficients are denoted by $U_S^{\rm K}$ and $U_T^{\rm K}$, respectively, then $U_T=|\vec q|^2U_T^{\rm K}$, $U_S^{\rm K}=-U_T$.
Note that the kinematical prefactors differ from those of Ref.~\cite{Kaiser:1997mw} due to a different sign convention for the potential $V$.}. Using a more consistent energy $E=m_N$ yields few-percent discrepancies in phase shifts at laboratory energies $T_\text{lab}$ near the pion-production threshold.

\begin{figure*}[tbp]
\centering
\includegraphics[width=0.96\textwidth]{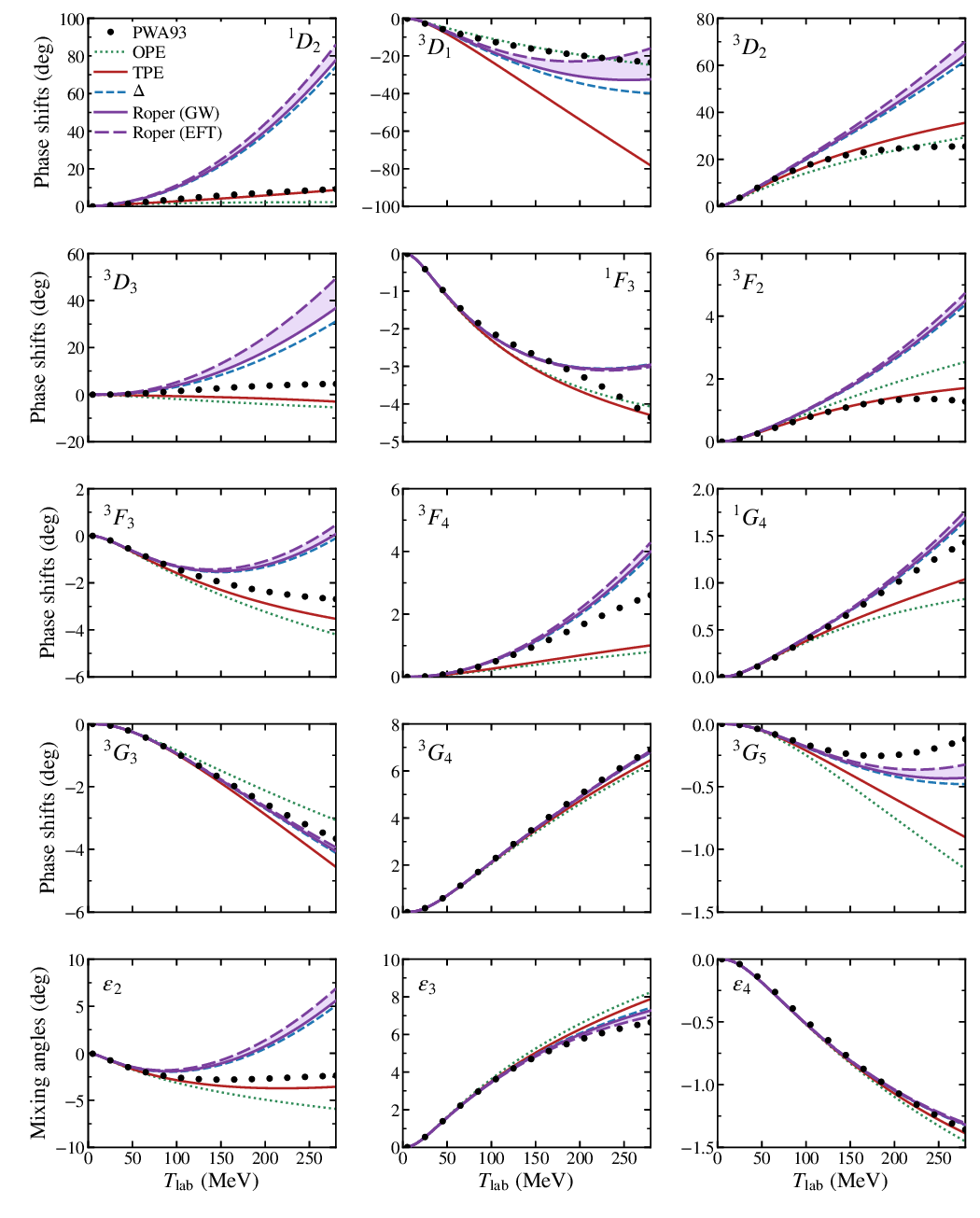}
\caption{
$D$-, $F$-, and $G$-wave phase shifts and the mixing angles $\varepsilon_{2,3,4}$, in degrees, as functions of the lab energy $T_{\rm lab}$, in MeV. The green dotted curves
represent OPE results, while the red solid curves are the results obtained after adding the leading TPE contributions with only nucleons in the intermediate states. The blue dashed curves show the corresponding results after including the $\Delta(1232)$ TPE contributions. The purple solid and dashed curves represent the results obtained by further including the full Roper contributions pure-Roper and mixed Delta-Roper sectors), with two sets of Roper parameters (``Roper (GW)" and ``Roper (EFT)"). The black dots denote the Nijmegen PSA~\cite{Stoks:1993tb}.
}
\label{fig:allphaseoverview}
\end{figure*}

$D$-, $F$-, and $G$-wave phase shifts as well as the mixing angles $\epsilon_{2,3,4}$ are shown in~\cref{fig:allphaseoverview} as a function of the laboratory energy $T_{\rm lab}=2p^2/m_N$ up to near the pion-production threshold ($\approx 280$ MeV). Results are shown for OPE~\cite{Kaiser:1997mw} and for successive additions to OPE: nucleon-only TPE~\cite{Kaiser:1997mw} (formally leading in Deltaless $\chi$EFT), TPE including Deltas, and TPE including further the Roper. They are compared with the Nijmegen phase-shift analysis (PSA)~\cite{Stoks:1993tb}.
In peripheral waves one expects~\cite{Nogga:2005hy,Birse:2005um,PavonValderrama:2016lqn,Wu:2018lai,Kaplan:2019znu,Peng:2020nyz} pions to be perturbative  thanks to the centrifugal barrier. The peripheral character of the $F$ and $G$ waves is manifest on the small phases and angles which are on the order of a few degrees. In contrast, the $D$-wave phase shifts  tend to be larger, indicating the increasing effects of the interactions.

As is well-known, OPE describes well the Nijmegen PSA results in most high partial waves. The long-range part of the formally leading nucleons TPE is, generically, a small effect that goes in the direction of the empirical phase shifts. In contrast, the long-range effects of the TPE with Deltas is relatively large, particularly in the lower phases and mixing angles. In Deltaless $\chi$EFT, the significance of the Delta is reflected in a formally subleading, but not smaller TPE.

Note that the curve labeled ``Delta'' in Fig.~\ref{fig:allphaseoverview} should not be expected to coincide numerically with the Deltaful phase shifts of Ref.~\cite{Krebs:2007rh}.  The present calculation uses dimensional regularization for the Delta TPE potential, whereas Ref.~\cite{Krebs:2007rh} employed spectral-function regularization (SFR) with a cutoff parameter $\tilde\Lambda=700$ MeV. This parameter suppresses the high-mass part of the two-pion spectral function and therefore removes part of the shorter-range component of the TPE potential, which is retained in dimensional regularization.  This difference is most visible in the $D$ waves, which are only marginally peripheral and
more sensitive to the shorter-range part of the interaction.  We have also checked, by varying the SFR cutoff, that the Delta contributions to the $D$-wave phase shifts show a noticeable cutoff dependence.  The discrepancy between the present Delta curve and the SFR result of Ref.~\cite{Krebs:2007rh} should therefore be understood mainly as a regularization effect in these low peripheral waves, rather than as a difference in the underlying long-range Delta-excitation mechanism.

The Delta generates the largest resonance-induced effects
in nearly every channel, particularly in the higher waves.
The additional Roper-dependent correction is smaller in all waves, with the parameter set ``N$^2$LO EFT'' giving the largest contribution. In most $F$ and $G$ waves the Roper curves lie close to the Delta curve, consistent with the stronger centrifugal suppression of the shorter-range Roper dynamics. The difference with the Delta curves is most pronounced in
the $D$ waves and in $\varepsilon_2$.  The difference between the two Roper parameter sets is also most pronounced in these channels.

The comparison with the Nijmegen PSA is strongly channel dependent.
The $^3G_4$ phase shift and $\varepsilon_4$ are comparatively insensitive to the resonance contributions over the displayed energy range.
The Delta and Roper-dependent corrections improve the description of $^1F_3$, $^1G_4$, $^3G_3$, $^3G_5$, and $\varepsilon_3$, whereas they overshoot the PSA in $^3F_2$, $^3F_3$, and $^3F_4$.
This is also true  in the lower waves $^1D_2$, $^3D_2$, and $^3D_3$, and they drive $\varepsilon_2$ through zero, away from the empirical trend. In contrast, in $^3D_1$, the Delta correction substantially reduces the excessive repulsion generated by leading TPE, and the Roper-dependent terms move the result further toward the empirical phase shift, especially for the ``N$^2$LO EFT" parameter set.

Thus the explicit resonances do not produce a universal improvement. Their impact is controlled by channel-dependent cancellations among the central and quasi-tensor components. For spin-singlet waves the relevant combination is $U_C+2U_T$. For spin-triplet waves, the mixing between $L=J-1$ and $L=J+1$ channels is generated by $U_T$, but both $U_C$ and $U_T$ enter in the phase shifts with channel-dependent angular projections.
The shifts in phase shifts therefore reflect the full isospin-projected central and quasi-tensor interactions after partial-wave projection, rather than any single momentum-space component alone.

The comparison with the PSA should be interpreted qualitatively.
The very reason nuclear bound states exist in $\chi$EFT, which is perturbative otherwise, is an infrared enhancement of nucleon-reducible diagrams over the nucleon-irreducible diagrams that make up the potential~\cite{Weinberg:1991um}. In lower waves this amounts to an expansion in $m_N Q V/4\pi$~\cite{Bedaque:2002mn, Hammer:2019poc}. For OPE, $V={\cal O}(f_\pi^{-2})={\cal O}(4\pi/m_N f_\pi)$ and power counting suggests the need for full iteration and the appearance of $S$-matrix poles at energies in the right ballpark, namely $\sim f_\pi/4\pi\sim 10$ MeV. In higher waves, factors of angular momentum suppress iterations~\cite{Nogga:2005hy,Birse:2005um,PavonValderrama:2016lqn,Kaplan:2019znu} and, for sufficiently large $L$, render pions perturbative. The transition from one regime to another is gradual and the optimal choice is still unsettled in the literature --- for a review, see Ref.~\cite{vanKolck:2020llt}. In any case, for a meaningful comparison with data at intermediate values of $L$, finite iterations of OPE need to be included together with the short-range interactions required for renormalization. For example, a recent implementation in Deltaless $\chi$EFT~\cite{Thim:2024yks} treats OPE and leading TPE effects as ${\cal O}(Q/M_{\rm QCD})$ and ${\cal O}(Q^3/M_{\rm QCD}^3)$, respectively, in uncoupled $L\ge 2$ waves. At $Q\sim f_\pi$, the infrared enhancement would then suggest including once-~\cite{Kaiser:1997mw} and twice-iterated OPE at the same time as leading TPE. In addition, some of the channels shown here are coupled to lower partial waves that are known to be nonperturbative, such as the coupling of $^3D_1$ to $^3S_1$ ~\cite{Fleming:1999ee,Nogga:2005hy,Birse:2005um,Wu:2018lai,Kaplan:2019znu} and the connection of $\varepsilon_2$ and $^3F_2$ to $^3P_2$ ~\cite{Peng:2020nyz}. Remaining discrepancies with the PSA, which should increase with energy, would reflect omitted higher-order effects, such as additional long-range interactions, recoil corrections, and contact interactions that contribute to peripheral waves.

The curves in Fig.~\ref{fig:allphaseoverview} should therefore be viewed as a decomposition of leading long-range resonance effects, rather than as a precision description of NN scattering. One can see them as a corroboration of the importance of Delta effects tied to its relatively small mass and significant pion-driven transition to the nucleon.
The pure-Delta contribution serves mainly as a benchmark for the size of explicit-resonance effects. The Roper contributes less, at least at physical pion mass, due to its larger mass and relatively weaker pion transition to the nucleon.

\section{Summary and outlook}
\label{SecOutlook}
We have derived the leading two-pion-exchange NN potentials generated by explicit Delta and Roper resonances in HB$\chi$PT.  The calculation includes the pure one- and two-Delta diagrams, the pure one- and two-Roper diagrams, and the mixed Delta--Roper box diagrams.  The results are given in a common momentum-space representation in dimensional regularization.  The analytic polynomial pieces are not unique and can be absorbed into short-range NN operators, while the nonanalytic parts encode the finite-range TPE dynamics. Our results for pure-Delta contributions agree with previous calculations. The additional long-range contributions from the Roper reduce to known subleading potentials in Deltaful $\chi$EFT when the Roper is integrated out.

The calculation shows a clear hierarchy among the explicit resonance contributions. The Delta gives the dominant correction to the long-range potential, as expected from its lower excitation energy and strong coupling to the pion-nucleon system. The Roper potential is smaller, the pure-Roper and mixed Delta-Roper contributions interfering constructively or destructively depending on the spin-isospin configuration. Roper effects tend to be larger in the isoscalar central and isovector quasi-tensor components. The magnitude of the effects is, however, strongly dependent on how Roper parameters are determined at the order of the calculation, as exemplified by two such sets. For one of these sets, the Roper contribution is a significant fraction of Delta's for the quasi-tensor component of the potential. Integrating out the Roper would miss this contribution at this order.

The same pattern is reflected in the perturbative phase shifts. The Delta produces the largest resonance-induced change, while the complete Roper-dependent correction, consisting of the pure-Roper and mixed Delta-Roper pieces, is concentrated mainly in the $D$ waves and in $\varepsilon_2$.  In most $F$ and $G$ waves the Roper-dependent effect is strongly suppressed.  The impact of the explicit resonances is nevertheless channel dependent: some partial waves move closer to the Nijmegen PSA, whereas others are overcorrected, indicating that the $D$ waves are only marginally peripheral and should be interpreted with caution in first-order perturbation theory.

The present work completes the leading long-range TPE contributions from the two lowest baryon resonances and their mixed excitation.  The hierarchy of resonance contributions at this order means that effects of an explicit Roper are marginal, suggesting that no further baryon resonances need to be included explicitly in the $\chi$EFT potential. A natural extension is to study higher-order corrections within the same long-range sector, such as recoil effects and other relativistic effects as well as subleading pion-baryon vertices, and to examine how stable the hierarchy found here remains beyond leading order.

\section*{Acknowledgments}
We thank Norbert Kaiser for useful comments and for pointing out an incorrect prefactor in the Roper diagrams in an earlier version of this work. This work is partly supported by the National Natural Science Foundation of China under Grant No.12435007, No.12505096,  the Chinese Postdoctoral Science Foundation under Grants No.2022M720360, and the Institute for Basic Science (IBS-R031-D1). This material is based upon work supported in part by the U.S. Department of Energy, Office of Science, Office of Nuclear Physics, under award DE-FG02-04ER41338.

\appendix
\section{Loop functions}
\label{app:loop-functions}

We collect here the loop functions that enter the resonance contributions in Sec.~\ref{SubSecDiag}.  We follow the tensor decomposition of Ref.~\cite{Wang:2018atz,Wang:2019ato}, with minor changes of notation. Dimensional regularization is used in $d$ spacetime dimensions, the scale it introduces being denoted by $\mu$.

The integrals appearing in two-pion exchange diagrams depend on the momentum transfer $q^\mu=(0,\vec{q})$ and involve masses that enter through propagators. In the following, we write $q$ as a shorthand for the four-vector $q^\mu=(0,\vec q)$, and every occurrence of $q^2$ means $q_\mu q^\mu=-\vec q^{\,2}$.
For compactness, we write the pion propagator in terms of
\begin{equation}
D_\pi(k)\equiv k^2-m^2+i0.
\label{eq:app-pion-denominator}
\end{equation}
The nucleon velocity is $v^\mu=(1,0)$ and the nucleon propagator is the inverse of
\begin{equation}
D_N(k,\omega)\equiv v\cdot k +\omega +i0.
\label{eq:app-nucleon-denominator}
\end{equation}

The triangle family of diagrams involves one heavy-baryon propagator and gives rise to the integrals
\begin{align}
\mathcal I_T^{\mu_1\cdots\mu_n}(m,\omega,q)
={}&i\mu^{4-d}
\int\!\frac{d^d\ell}{(2\pi)^d}
\frac{\ell^{\mu_1}\cdots\ell^{\mu_n}}
{\mathcal D_T(\ell;m,\omega,q)} ,
\label{eq:app-triangle-integral}
\\
\mathcal D_T(\ell;m,\omega,q)
={}&
D_N(\ell,\omega)D_\pi(\ell)D_\pi(\ell+q).
\label{eq:app-triangle-denominator}
\end{align}
The crossed-box ($X$) and planar-box ($B$) families, in turn, involve
\begin{align}
\mathcal I_Z^{\mu_1\cdots\mu_n}(m,\omega_1,\omega_2,q)
={}&i\mu^{4-d}
\int\!\frac{d^d\ell}{(2\pi)^d}
\frac{\ell^{\mu_1}\cdots\ell^{\mu_n}}
{\mathcal D_Z(\ell;m,\omega_1,\omega_2,q)},
\label{eq:app-box-integral}
\\
\mathcal D_Z(\ell;m,\omega_1,\omega_2,q)
={}&D_N(\ell,\omega_1) D_N(s_Z \ell, \omega_2)
D_\pi(\ell)D_\pi(\ell+q),
\label{eq:app-box-denominator}
\end{align}
where $Z=B,X$ and $s_X=+1$, $s_B=-1$.

The TPE potential is obtained from the tensor decompositions of the above integrals. In these decompositions we use curly brackets to denote the unnormalized sum over distinct permutations; for example,
\begin{align}
q^{\{\alpha}v^{\beta\}}&=q^\alpha v^\beta+q^\beta v^\alpha,
\\
g^{\{\alpha\beta}q^{\gamma\}}&=
g^{\alpha\beta}q^\gamma+g^{\alpha\gamma}q^\beta
+g^{\beta\gamma}q^\alpha.
\end{align}
The scalar coefficients $\mathcal J^Z_{ij}$, with $Z=T,B,X$, are fixed by the tensor decompositions
\begin{align}
\mathcal I_Z^{\alpha\beta}
={}&g^{\alpha\beta}\mathcal J^Z_{21}
+q^\alpha q^\beta\mathcal J^Z_{22}
+v^\alpha v^\beta\mathcal J^Z_{23}
+q^{\{\alpha}v^{\beta\}}\mathcal J^Z_{24},
\label{eq:app-box-rank2}
\\
\mathcal I_Z^{\alpha\beta\gamma}
={}&g^{\{\alpha\beta}q^{\gamma\}}\mathcal J^Z_{31}
+q^\alpha q^\beta q^\gamma\mathcal J^Z_{32}
+q^{\{\alpha}q^\beta v^{\gamma\}}\mathcal J^Z_{33}
+g^{\{\alpha\beta}v^{\gamma\}}\mathcal J^Z_{34}
\nonumber\\
&+q^{\{\alpha}v^\beta v^{\gamma\}}\mathcal J^Z_{35}
+v^\alpha v^\beta v^\gamma\mathcal J^Z_{36},
\label{eq:app-box-rank3}
\\
\mathcal I_Z^{\alpha\beta\gamma\delta}
={}&g^{\{\alpha\beta}g^{\gamma\delta\}}\mathcal J^Z_{41}
+g^{\{\alpha\beta}q^\gamma q^{\delta\}}\mathcal J^Z_{42}
+q^\alpha q^\beta q^\gamma q^\delta\mathcal J^Z_{43}
+g^{\{\alpha\beta}v^\gamma v^{\delta\}}\mathcal J^Z_{44}
\nonumber\\
&+v^\alpha v^\beta v^\gamma v^\delta\mathcal J^Z_{45}
+q^{\{\alpha}q^\beta q^\gamma v^{\delta\}}\mathcal J^Z_{46}
+q^{\{\alpha}q^\beta v^\gamma v^{\delta\}}\mathcal J^Z_{47}
\nonumber\\
&+q^{\{\alpha}v^\beta v^\gamma v^{\delta\}}\mathcal J^Z_{48}
+g^{\{\alpha\beta}q^\gamma v^{\delta\}}\mathcal J^Z_{49},
\label{eq:app-box-rank4}
\end{align}
where the arguments of the scalar functions are to be understood. The triangle functions appearing explicitly in TPE are $\mathcal J^T_{24,33,34}(m,\omega,q)$. The box functions needed are $\mathcal J^{B,X}_{21,22,31,32,41,42,43}(m,\omega_1, \omega_2,q)$.

The box functions can be obtained from the triangle functions. For the crossed-box family, partial fractioning the two heavy-baryon propagators gives, for any scalar coefficient $\mathcal J_{ij}$,
\begin{align}
\mathcal J_{ij}^X(m,\omega_1,\omega_2,q)
=
\begin{cases}
\displaystyle
-\frac{
\mathcal J_{ij}^T(m,\omega_2,q)
-\mathcal J_{ij}^T(m,\omega_1,q)}
{\omega_2-\omega_1},
& \omega_1\neq\omega_2,
\\[3mm]
\displaystyle
-\left.
\frac{\partial}{\partial\omega}
\mathcal J_{ij}^T(m,\omega,q)
\right|_{\omega=\omega_1},
& \omega_1=\omega_2.
\end{cases}
\label{eq:app-JR-from-JT}
\end{align}
For the planar-box family, if $\mathcal J_{ij}^S(m,\omega,q)$ denotes the triangle coefficient obtained by replacing the heavy denominator $v\cdot\ell+\omega+i0$ with $-v\cdot\ell+\omega+i0$, partial fractioning gives
\begin{equation}
\mathcal J_{ij}^B(m,\omega_1,\omega_2,q)
=\frac{
\mathcal J_{ij}^T(m,\omega_1,q)
+\mathcal J_{ij}^S(m,\omega_2,q)}
{\omega_1+\omega_2}.
\label{eq:app-JB-from-JTS}
\end{equation}
In the static kinematics used here, $v\cdot q=0$, and the two pion masses are equal.  The $S$- and $T$-family coefficients needed in the main text then obey
\begin{equation}
\mathcal J_{ij}^S(m,\omega,q)=\mathcal J_{ij}^T(m,\omega,q),
\qquad
ij=21,22,31,32,41,42,43,
\label{eq:app-JS-equals-JT}
\end{equation}
and the required planar-box functions are
\begin{align}
\mathcal J_{ij}^B(m,\omega_1,\omega_2,q)
={}&\frac{
\mathcal J_{ij}^T(m,\omega_1,q)
+\mathcal J_{ij}^T(m,\omega_2,q)}
{\omega_1+\omega_2},
\qquad ij=21,22,31,32,41,42,43.
\label{eq:app-JB-from-JT}
\end{align}
All planar-box arguments occurring in Sec.~\ref{SubSecDiag} obey $\omega_1+\omega_2\neq0$.  Thus, unlike the special case $\omega_1=\omega_2=0$, they do not generate the two-nucleon-reducible pinch singularity.

We separate the ultraviolet poles by writing
\begin{equation}
\mathcal J_{ij}^{Z,(d)}
=\frac{r_{ij}^{Z}}{d-4}+\mathcal J_{ij}^{Z}
+\mathcal O(d-4),
\qquad
r_{ij}^{Z}\equiv\lim_{d\to4}(d-4)\mathcal J_{ij}^{Z,(d)},
\qquad Z=T,B,X .
\label{eq:app-J-Laurent}
\end{equation}

All the required triangle functions can be written as parameter integrals. Defining the dimensionless functions
\begin{align}
A(x)
&=
x(x-1)\frac{q^2}{m^2}
-\frac{\omega^2}{m^2}
+1 -i0^+,
\\
\Delta(x,y)
&=
\frac{\omega^2}{m^2}y^2+A(x),
\label{eq:app-A-Delta}
\\
L(\mu)
&\equiv
\frac{1}{32\pi^2}
\left(
\gamma_E+\ln\frac{m^2}{4\pi\mu^2}
\right),
\label{eq:app-Lmu}
\end{align}
one finds~\footnote{For compactness, the common arguments $(m,\omega,q)$ are suppressed.}
\begin{align}
\mathcal J^T_{21}
={}&
2\omega L(\mu)
+\frac{\omega}{16\pi^2}
\int_0^1\!dx\int_{-1}^0\!dy\,\ln\Delta(x,y)
+\frac{m}{16\pi}\int_0^1\!dx\,A^{1/2}(x),
\label{eq:app-JT21}
\\
\mathcal J^T_{22}
={}&
\frac{\omega}{8\pi^2m^2}
\int_0^1\!dx\int_{-1}^0\!dy\,
\frac{x^2}{\Delta(x,y)}
+\frac{1}{16\pi m}\int_0^1\!dx\,x^2A^{-1/2}(x),
\label{eq:app-JT22}
\\
\mathcal J^T_{24}
={}&
-L(\mu)
+\frac{\omega^2}{8\pi^2m^2}
\int_0^1\!dx\int_{-1}^0\!dy\,
\frac{x(y+1)}{\Delta(x,y)}
-\frac{1}{16\pi^2}\int_0^1\!dx\,x\ln A(x)
\nonumber\\
{}&+\frac{\omega}{16\pi m}\int_0^1\!dx\,xA^{-1/2}(x).
\label{eq:app-JT24}
\end{align}
for the rank-two tensors.
\begin{align}
\mathcal J^T_{31}
={}&
-\omega L(\mu)
-\frac{\omega}{16\pi^2}
\int_0^1\!dx\int_{-1}^0\!dy\,x\ln\Delta(x,y)
-\frac{m}{16\pi}\int_0^1\!dx\,xA^{1/2}(x),
\label{eq:app-JT31}
\\
\mathcal J^T_{32}
={}&
-\frac{\omega}{8\pi^2m^2}
\int_0^1\!dx\int_{-1}^0\!dy\,
\frac{x^3}{\Delta(x,y)}
-\frac{1}{16\pi m}\int_0^1\!dx\,x^3A^{-1/2}(x),
\label{eq:app-JT32}
\\
\mathcal J^T_{33}
={}&
\frac{2}{3}L(\mu)
-\frac{\omega^2}{8\pi^2m^2}
\int_0^1\!dx\int_{-1}^0\!dy\,
\frac{x^2(y+1)}{\Delta(x,y)}
+\frac{1}{16\pi^2}\int_0^1\!dx\,x^2\ln A(x)
\nonumber\\
{}&-\frac{\omega}{16\pi m}\int_0^1\!dx\,x^2A^{-1/2}(x),
\label{eq:app-JT33}
\\
\mathcal J^T_{34}
={}&
\left(m^2-\frac{q^2}{6}-2\omega^2\right)
L(\mu)
-\frac{1}{32\pi^2}\left(m^2-\frac{q^2}{6}-\omega^2\right)
-\frac{\omega^2}{16\pi^2}
\int_0^1\!dx\int_{-1}^0\!dy\,
(y+1)\ln\Delta(x,y)
\nonumber\\
&-\frac{\omega m}{16\pi}\int_0^1\!dx\,A^{1/2}(x)
+\frac{m^2}{32\pi^2}
\int_0^1\!dx\,A(x)\ln A(x).
\label{eq:app-JT34}
\end{align}
for the rank-three tensors,
\begin{align}
\mathcal J^T_{41}
={}&
\omega\left(m^2-\frac{q^2}{6}-\frac{2\omega^2}{3}\right)
\left[L(\mu)-\frac{1}{32\pi^2}\right]
+\frac{\omega m^2}{32\pi^2}
\int_0^1\!dx\int_{-1}^0\!dy\,
\Delta(x,y)\ln\Delta(x,y)
\nonumber\\
{}&+\frac{m^3}{48\pi}\int_0^1\!dx\,A(x)^{3/2},
\label{eq:app-JT41}
\\
\mathcal J^T_{42}
={}&
\frac{2\omega}{3}L(\mu)
+\frac{\omega}{16\pi^2}
\int_0^1\!dx\int_{-1}^0\!dy\,
x^2\ln\Delta(x,y)
+\frac{m}{16\pi}\int_0^1\!dx\,x^2A^{1/2}(x),
\label{eq:app-JT42}
\\
\mathcal J^T_{43}
={}&
\frac{\omega}{8\pi^2m^2}
\int_0^1\!dx\int_{-1}^0\!dy\,
\frac{x^4}{\Delta(x,y)}
+\frac{1}{16\pi m}\int_0^1\!dx\,x^4A^{-1/2}(x).
\label{eq:app-JT43}
\end{align}
for the rank-four tensors.

The residues $r_{ij}^Z$ involve
\begin{equation}
\kappa\equiv\frac{1}{16\pi^2}.
\end{equation}
For the triangle family the nonvanishing residues are
\begin{align}
r_{21}^{T}={}&2\kappa\omega,
&
r_{24}^{T}={}&-\kappa,
&
r_{31}^{T}={}&-\kappa\omega,
&
r_{33}^{T}={}&\frac{2\kappa}{3},
\nonumber\\
r_{34}^{T}={}&
\kappa\left(m^2-\frac{q^2}{6}-2\omega^2\right),
&
r_{41}^{T}={}&
\kappa\omega\left(m^2-\frac{q^2}{6}-\frac{2\omega^2}{3}\right),
&
r_{42}^{T}={}&\frac{2\kappa\omega}{3},
\label{eq:app-J-residues}
\end{align}
with $r_{22}^{T}=r_{32}^{T}=r_{43}^{T}=0$. Applying the partial-fraction relations in Eqs.~\eqref{eq:app-JR-from-JT} and~\eqref{eq:app-JB-from-JT} gives
\begin{align}
r_{21}^{B,X}=\pm2\kappa,\quad
r_{31}^{B,X}=\mp\kappa,\quad
r_{42}^{B,X}=\pm\frac{2\kappa}{3},\quad
r_{41}^{B,X}=
\pm\kappa\left[
m^2-\frac{q^2}{6}
-\frac{2}{3}\left(
\omega_1^2\mp\omega_1\omega_2+\omega_2^2
\right)
\right],
\label{eq:app-box-residues}
\end{align}
Here and below, the upper (lower) sign corresponds to $B$ ($X$). The remaining residues vanish, $r_{22}^{B,X}=r_{32}^{B,X}=r_{43}^{B,X}=0$.

TPE generates certain combinations of the scalar functions with $d$-dependent coefficients. Expanding the dimension-dependent coefficients through first order in $d-4$ and using the Laurent decomposition in
Eq.~\eqref{eq:app-J-Laurent} generates finite contributions proportional to the pole residues.

The quasi-tensor components arise solely from box diagrams and involve only $q^2 \mathcal J_{21}^{B,X}(m_\pi,\omega_1,\omega_2,q)$, with $\omega_{1,2}= 0, -\delta, -\rho$.
The relevant combinations for TPE are
\begin{align}
\mathcal K_\pm^{RN}(q)
\equiv{}&
\mathcal J_{21}^{B}(m_\pi,-\rho,0,q)
\pm
\mathcal J_{21}^{X}(m_\pi,-\rho,0,q),
\nonumber\\
\mathcal K_\pm^{RR}(q)
\equiv{}&
\mathcal J_{21}^{B}(m_\pi,-\rho,-\rho,q)
\pm
\mathcal J_{21}^{X}(m_\pi,-\rho,-\rho,q),
\nonumber\\
\mathcal K_+^{\Delta N}(q)
\equiv{}&
\frac{1}{3}
\left[
\mathcal J_{21}^{B}(m_\pi,-\delta,0,q)
+
\mathcal J_{21}^{X}(m_\pi,-\delta,0,q)
\right],
\nonumber\\
\mathcal K_-^{\Delta N}(q)
\equiv{}&
\frac{1}{3}
\left[
\mathcal J_{21}^{B}(m_\pi,-\delta,0,q)
-
\mathcal J_{21}^{X}(m_\pi,-\delta,0,q)
\right]
-\frac{4\kappa}{9},
\nonumber\\
\mathcal K_+^{\Delta R}(q)
\equiv{}&
\frac{1}{3}
\left[
\mathcal J_{21}^{B}(m_\pi,-\delta,-\rho,q)
+
\mathcal J_{21}^{X}(m_\pi,-\delta,-\rho,q)
\right],
\nonumber\\
\mathcal K_-^{\Delta R}(q)
\equiv{}&
\frac{1}{3}
\left[
\mathcal J_{21}^{B}(m_\pi,-\delta,-\rho,q)
-
\mathcal J_{21}^{X}(m_\pi,-\delta,-\rho,q)
\right]
-\frac{4\kappa}{9},
\nonumber\\
\mathcal K_+^{\Delta\Delta}(q)
\equiv{}&
\frac{1}{9}
\left[
\mathcal J_{21}^{B}(m_\pi,-\delta,-\delta,q)
+
\mathcal J_{21}^{X}(m_\pi,-\delta,-\delta,q)
\right],
\nonumber\\
\mathcal K_-^{\Delta\Delta}(q)
\equiv{}&
\frac{1}{9}
\left[
\mathcal J_{21}^{B}(m_\pi,-\delta,-\delta,q)
-
\mathcal J_{21}^{X}(m_\pi,-\delta,-\delta,q)
\right]
-\frac{8\kappa}{27}.
\label{eq:K-kernels-main}
\end{align}

The central components emerge not only from boxes but also from triangle diagrams and involve several scalar functions.
If we define
\begin{align}
\mathcal T[\mathcal J^{T}]
\equiv{}&3\mathcal J_{34}^{T}
+q^2\left(\mathcal J_{24}^{T}+\mathcal J_{33}^{T}\right)
+\kappa\left(m^2-\frac{q^2}{6}-2\omega^2\right),
\label{eq:T-kernel-main}
\\
\mathcal S[\mathcal J^{T}]
\equiv{}&2\mathcal J_{34}^{T}
+\frac{2q^2}{3}\left(\mathcal J_{24}^{T}+\mathcal J_{33}^{T}\right)
+\kappa\left(m^2-\frac{11q^2}{54}-2\omega^2\right),
\label{eq:S-kernel-main}
\end{align}
then the triangle contributions are expressed through the abbreviations
\begin{align*}
\mathcal S^\Delta(q)\equiv{}&\mathcal S[\mathcal J^{T}(m_\pi,-\delta,q)],
\\
\mathcal T^R(q)\equiv{}&\mathcal T[\mathcal J^{T}(m_\pi,-\rho,q)].
\end{align*}
For the box family, whose common arguments are $(m,\omega_1,\omega_2,q)$,  with the upper (lower) signs corresponding to $B$ ($X$), we define
\begin{align}
\mathcal R[\mathcal J^{B,X}]
\equiv{}&
15\mathcal J_{41}^{B,X}
+10q^2\left(
\mathcal J_{42}^{B,X}+\mathcal J_{31}^{B,X}
\right)
+q^2\mathcal J_{21}^{B,X}
+q^4\left(
\mathcal J_{43}^{B,X}
+2\mathcal J_{32}^{B,X}
+\mathcal J_{22}^{B,X}
\right)
\nonumber\\
&\pm\kappa\left[
8m^2-2q^2
-\frac{16}{3}
\left(
\omega_1^2\mp\omega_1\omega_2+\omega_2^2
\right)
\right],
\label{eq:R-kernel-main}
\\
\mathcal P_{B,X}
\equiv{}&
\pm\kappa\left[
15m^2-\frac{23q^2}{6}
-10\left(
\omega_1^2\mp\omega_1\omega_2+\omega_2^2
\right)
\right],
\label{eq:P-kernel-main}
\\
\mathcal C[\mathcal J^{Z}]
\equiv{}&
\frac{2}{3}\mathcal R[\mathcal J^{Z}]
+\frac{1}{9}\mathcal P_Z,
\label{eq:C-kernel-main}
\\
\mathcal D[\mathcal J^{Z}]
\equiv{}&
\frac{4}{9}\mathcal R[\mathcal J^{Z}]
+\frac{4}{27}\mathcal P_Z,
\qquad Z=B,X,
\label{eq:D-kernel-main}
\end{align}
so that the box diagrams are given in terms of
\begin{align*}
\mathcal C_{B,X}^{\Delta N}(q)&\equiv
\mathcal C\!\left[\mathcal J^{B,X}(m_\pi,-\delta,0,q)\right],
&\mathcal D_{B,X}^{\Delta\Delta}(q)&\equiv
\mathcal D\!\left[\mathcal J^{B,X}(m_\pi,-\delta,-\delta,q)\right],
\\
\mathcal R_{B,X}^{RN}(q)&\equiv
\mathcal R\!\left[\mathcal J^{B,X}(m_\pi,-\rho,0,q)\right],
&\mathcal C_{B,X}^{\Delta R}(q)&\equiv
\mathcal C\!\left[\mathcal J^{B,X}(m_\pi,-\delta,-\rho,q)\right],
\\
\mathcal R_{B,X}^{RR}(q)&\equiv
\mathcal R\!\left[\mathcal J^{B,X}(m_\pi,-\rho,-\rho,q)\right].
\end{align*}

The functions $\mathcal S^{\Delta}$, $\mathcal T^R$, $\mathcal K^{\Delta N, \Delta\Delta, RN, RR, \Delta R}_{\pm}$, $\mathcal C^{\Delta N,\Delta R}_{B,X}$, $\mathcal R^{RN,RR}_{B,X}$, and $\mathcal D^{\Delta\Delta}_{B,X}$ described above give the TPE in Eq.~\eqref{eq:mixed-total-main}.
\section{Polynomial-subtraction parameters}
\label{app:polynomial-subtractions}

The subtraction coefficients $a_0$ and $a_1$ used in Eqs.~\eqref{eq:poly-subtraction} and~\eqref{eq:poly-coefficients} of the main text are collected here. The renormalization scale $\mu=1$ GeV is taken for concreteness.

Table~\ref{tab:poly-nn-delta} gives the subtraction coefficients extracted from the leading nucleon-only TPE from Ref.~\cite{Kaiser:1997mw}. It also lists the coefficients obtained from Eq.~\eqref{eq:delta-total-main} of the main text for the pure-Delta contribution.
The coefficients extracted from the pure-Roper TPE in Eq.~\eqref{eq:roper-total-main} and from the mixed Delta-Roper potentials in Eq.~\eqref{eq:mixed-total-main} are given, respectively, in Tables~\ref{tab:poly-roper} and~\ref{tab:poly-mixed}. These tables present results with the two sets of Roper parameters we consider in this manuscript, labeled ``N$^2$LO GW" and ``N$^2$LO EFT", as discussed in Sec.~\ref{SecLECs}.
Finally, the subtraction coefficients for the combined integrated-out numerator and combined explicit-Roper denominator entering Eq.~\eqref{eq:roper-integrated-out-ratio} are given in Table~\ref{tab:poly-integrated-out-ratio}.

\begin{table}[H]
\caption{Polynomial coefficients for the leading nucleon-only TPE and pure-Delta contributions.}
\label{tab:poly-nn-delta}
\begin{ruledtabular}
\begin{tabular}{c cc cc}
Component & \multicolumn{2}{c}{Nucleons only} & \multicolumn{2}{c}{Delta}\\
& $a_0\, [\mathrm{GeV}^{-2}]$
& $a_1\, [\mathrm{GeV}^{-4}]$
& $a_0\, [\mathrm{GeV}^{-2}]$
& $a_1\, [\mathrm{GeV}^{-4}]$
\\
\hline
$V_C$ & $0$ & $0$ & $-1.20\times10^{2}$ & $-2.32\times10^{3}$\\
$W_C$ & $1.74\times10^{1}$ & $1.37\times10^{2}$ & $-3.48\times10^{-1}$ & $-4.00\times10^{2}$\\
$V_T$ & $0$ & $2.66\times10^{2}$ & $0$ & $-3.50\times10^{2}$\\
$W_T$ & $0$ & $0$ & $0$ & $-1.24\times10^{2}$\\
\end{tabular}
\end{ruledtabular}
\end{table}

\begin{table}[H]
\caption{Polynomial coefficients for the pure-Roper TPE contribution with the ``N$^2$LO GW'' and ``N$^2$LO EFT'' parameter sets.}
\label{tab:poly-roper}
\begin{ruledtabular}
\begin{tabular}{c cc cc}
Component & \multicolumn{2}{c}{Roper (GW)} & \multicolumn{2}{c}{Roper (EFT)}\\
& $a_0\, [\mathrm{GeV}^{-2}]$
& $a_1\, [\mathrm{GeV}^{-4}]$
& $a_0\, [\mathrm{GeV}^{-2}]$
& $a_1\, [\mathrm{GeV}^{-4}]$
\\
\hline
$V_C$ & $-3.82\times10^{-1}$ & $-5.41\times10^{1}$ & $8.85\times10^{1}$ & $-2.34\times10^{2}$\\
$W_C$ & $-1.10\times10^{1}$ & $1.94$ & $4.68\times10^{1}$ & $-1.34\times10^{2}$\\
$V_T$ & $0$ & $3.04\times10^{1}$ & $0$ & $3.16\times10^{1}$\\
$W_T$ & $0$ & $-1.88\times10^{1}$ & $0$ & $-8.14\times10^{1}$\\
\end{tabular}
\end{ruledtabular}
\end{table}

\begin{table}[H]
\caption{Polynomial coefficients for the mixed Delta--Roper TPE contribution with the ``N$^2$LO GW'' and ``N$^2$LO EFT'' parameter sets.}
\label{tab:poly-mixed}
\begin{ruledtabular}
\begin{tabular}{c cc cc}
Component & \multicolumn{2}{c}{Delta--Roper (GW)} & \multicolumn{2}{c}{Delta--Roper (EFT)}\\
& $a_0\, [\mathrm{GeV}^{-2}]$
& $a_1\, [\mathrm{GeV}^{-4}]$
& $a_0\, [\mathrm{GeV}^{-2}]$
& $a_1\, [\mathrm{GeV}^{-4}]$
\\
\hline
$V_C$ & $1.60\times10^{1}$ & $-1.59\times10^{2}$ & $1.86\times10^{2}$ & $-5.68\times10^{2}$\\
$W_C$ & $1.55$ & $4.79\times10^{1}$ & $-9.74\times10^{1}$ & $2.44\times10^{2}$\\
$V_T$ & $0$ & $-3.61\times10^{1}$ & $0$ & $-9.22\times10^{1}$\\
$W_T$ & $0$ & $-1.38\times10^{1}$ & $0$ & $-4.94\times10^{1}$\\
\end{tabular}
\end{ruledtabular}
\end{table}

\begin{table}[H]
\caption{Polynomial coefficients for the combined integrated-out numerator,
$
Z^{(1\slashed R)}+Z^{(\Delta\slashed R)}$, and the combined explicit-Roper denominator, $
Z^{(R)}+Z^{(\Delta R)}$, entering Eq.~\eqref{eq:roper-integrated-out-ratio}.
For the numerator, the integrated-out mixed Delta--Roper contribution is evaluated using only the displayed terms in Eq.~\eqref{eq:delta-roper-integrated-out}.}
\label{tab:poly-integrated-out-ratio}
\begin{ruledtabular}
\begin{tabular}{
c cc cc}
Component
& \multicolumn{2}{c}{N$^2$LO GW}
& \multicolumn{2}{c}{N$^2$LO EFT}
\\
& $a_0\,[\mathrm{GeV}^{-2}]$
& $a_1\,[\mathrm{GeV}^{-4}]$
& $a_0\,[\mathrm{GeV}^{-2}]$
& $a_1\,[\mathrm{GeV}^{-4}]$
\\
\hline
$V_C^{(1\slashed R)}+V_C^{(\Delta\slashed R)}$
& $-1.47\times10^{1}$  & $1.22\times10^{2}$
& $-4.53\times10^{1}$  & $3.75\times10^{2}$
\\
$V_C^{(R)}+V_C^{(\Delta R)}$
& $1.56\times10^{1}$  & $-2.13\times10^{2}$
& $2.75\times10^{2}$ & $-8.02\times10^{2}$
\\
$W_T^{(1\slashed R)}+W_T^{(\Delta\slashed R)}$
& $0$ & $-1.66$
& $0$ & $-5.08$
\\
$W_T^{(R)}+W_T^{(\Delta R)}$
& $0$ & $-3.27\times10^{1}$
& $0$ & $-1.31\times10^{2}$
\end{tabular}
\end{ruledtabular}
\end{table}
\bibliography{refs}

@article{Bedaque:2002mn,
    author = "Bedaque, Paulo F. and van Kolck, Ubirajara",
    title = "{Effective field theory for few nucleon systems}",
    eprint = "nucl-th/0203055",
    archivePrefix = "arXiv",
    doi = "10.1146/annurev.nucl.52.050102.090637",
    journal = "Ann. Rev. Nucl. Part. Sci.",
    volume = "52",
    pages = "339--396",
    year = "2002"
}

@article{Thim:2024yks,
    author = {Thim, Oliver and Ekstr{\"o}m, Andreas and Forss{\'e}n, Christian},
    title = "{Perturbative computations of neutron-proton scattering observables using renormalization-group invariant chiral effective field theory up to N3LO}",
    eprint = "2402.15325",
    archivePrefix = "arXiv",
    primaryClass = "nucl-th",
    doi = "10.1103/PhysRevC.109.064001",
    journal = "Phys. Rev. C",
    volume = "109",
    number = "6",
    pages = "064001",
    year = "2024"
}

@article{vanKolck:2020llt,
    author = "van Kolck, U.",
    title = "{The Problem of Renormalization of Chiral Nuclear Forces}",
    eprint = "2003.06721",
    archivePrefix = "arXiv",
    primaryClass = "nucl-th",
    doi = "10.3389/fphy.2020.00079",
    journal = "Front. in Phys.",
    volume = "8",
    pages = "79",
    year = "2020"
}

@article{Lu:2025ubc,
    author = "Lu, Jun-Xu and Geng, Li-Sheng",
    title = "{Two-pion exchange contributions to the relativistic chiral nuclear force at N$^3$LO}",
    eprint = "2511.18023",
    archivePrefix = "arXiv",
    primaryClass = "nucl-th",
    month = "11",
    year = "2025"
}

@article{Xiao:2025ufy,
    author = "Xiao, Yang and Lu, Jun-Xu and Geng, Li-Sheng",
    title = "{Charge dependent nucleon-nucleon potentials in covariant chiral effective field theory}",
    eprint = "2504.15598",
    archivePrefix = "arXiv",
    primaryClass = "nucl-th",
    month = "4",
    year = "2025"
}

@article{Lu:2025syk,
    author = "Lu, Jun-Xu and Xiao, Yang and Liu, Zhi-Wei and Geng, Li-Sheng",
    title = "{Relativistic chiral nuclear forces: Status and prospects}",
    eprint = "2501.17185",
    archivePrefix = "arXiv",
    primaryClass = "nucl-th",
    doi = "10.1142/S0218301325430074",
    journal = "Int. J. Mod. Phys. E",
    volume = "34",
    number = "11",
    pages = "2543007",
    year = "2025"
}

@article{Xiao:2024jmu,
    author = "Xiao, Yang and Lu, Jun-Xu and Geng, Li-Sheng",
    title = "{Reexamination of antinucleon-nucleon interactions in covariant chiral effective field theory}",
    eprint = "2406.01292",
    archivePrefix = "arXiv",
    primaryClass = "nucl-th",
    doi = "10.1103/PhysRevC.110.064002",
    journal = "Phys. Rev. C",
    volume = "110",
    number = "6",
    pages = "064002",
    year = "2024"
}

@article{Weinberg:1968de,
    author = "Weinberg, Steven",
    title = "{Nonlinear realizations of chiral symmetry}",
    doi = "10.1103/PhysRev.166.1568",
    journal = "Phys. Rev.",
    volume = "166",
    pages = "1568--1577",
    year = "1968"
}

@article{Coleman:1969sm,
    author = "Coleman, Sidney R. and Wess, J. and Zumino, Bruno",
    title = "{Structure of phenomenological Lagrangians. 1.}",
    doi = "10.1103/PhysRev.177.2239",
    journal = "Phys. Rev.",
    volume = "177",
    pages = "2239--2247",
    year = "1969"
}

@article{Callan:1969sn,
    author = "Callan, Jr., Curtis G. and Coleman, Sidney R. and Wess, J. and Zumino, Bruno",
    title = "{Structure of phenomenological Lagrangians. 2.}",
    doi = "10.1103/PhysRev.177.2247",
    journal = "Phys. Rev.",
    volume = "177",
    pages = "2247--2250",
    year = "1969"
}

@article{Weinberg:1990rz,
    author = "Weinberg, Steven",
    title = "{Nuclear forces from chiral Lagrangians}",
    reportNumber = "UTTG-31-90",
    doi = "10.1016/0370-2693(90)90938-3",
    journal = "Phys. Lett. B",
    volume = "251",
    pages = "288--292",
    year = "1990"
}

@article{Weinberg:1991um,
    author = "Weinberg, Steven",
    title = "{Effective chiral Lagrangians for nucleon - pion interactions and nuclear forces}",
    reportNumber = "UTTG-03-91",
    doi = "10.1016/0550-3213(91)90231-L",
    journal = "Nucl. Phys. B",
    volume = "363",
    pages = "3--18",
    year = "1991"
}

@article{Ordonez:1992xp,
    author = "Ord\'o\~nez, C. and van Kolck, U.",
    title = "{Chiral lagrangians and nuclear forces}",
    reportNumber = "UTTG-01-92",
    doi = "10.1016/0370-2693(92)91404-W",
    journal = "Phys. Lett. B",
    volume = "291",
    pages = "459--464",
    year = "1992"
}

@article{Weinberg:1992yk,
    author = "Weinberg, Steven",
    title = "{Three body interactions among nucleons and pions}",
    eprint = "hep-ph/9209257",
    archivePrefix = "arXiv",
    reportNumber = "UTTG-11-92",
    doi = "10.1016/0370-2693(92)90099-P",
    journal = "Phys. Lett. B",
    volume = "295",
    pages = "114--121",
    year = "1992"
}

@article{Ordonez:1993tn,
    author = "Ord\'o\~nez, C. and Ray, L. and van Kolck, U.",
    title = "{Nucleon-nucleon potential from an effective chiral Lagrangian}",
    reportNumber = "UTTG-31-93, DOE-ER-40427-23-N93",
    doi = "10.1103/PhysRevLett.72.1982",
    journal = "Phys. Rev. Lett.",
    volume = "72",
    pages = "1982--1985",
    year = "1994"
}

@article{vanKolck:1994yi,
    author = "van Kolck, U.",
    title = "{Few nucleon forces from chiral Lagrangians}",
    doi = "10.1103/PhysRevC.49.2932",
    journal = "Phys. Rev. C",
    volume = "49",
    pages = "2932--2941",
    year = "1994"
}

@article{Ordonez:1995rz,
    author = "Ord\'o\~nez, C. and Ray, L. and van Kolck, U.",
    title = "{The Two nucleon potential from chiral Lagrangians}",
    eprint = "hep-ph/9511380",
    archivePrefix = "arXiv",
    reportNumber = "UTTG-15-95",
    doi = "10.1103/PhysRevC.53.2086",
    journal = "Phys. Rev. C",
    volume = "53",
    pages = "2086--2105",
    year = "1996"
}

@article{Weinberg:1978kz,
    author = "Weinberg, Steven",
    editor = "Deser, S.",
    title = "{Phenomenological Lagrangians}",
    reportNumber = "HUTP-78-A051A",
    doi = "10.1016/0378-4371(79)90223-1",
    journal = "Physica A",
    volume = "96",
    number = "1-2",
    pages = "327--340",
    year = "1979"
}

@article{Valderrama:2010aw,
    author = "Pav\'on Valderrama, Manuel",
    editor = "Nieves, Juan M. and Oset, Eulogio and Vicente Vacas, Manuel J.",
    title = "{Perturbative Renormalizability of Chiral Two Pion Exchange and Power Counting in Nucleon-Nucleon Scattering}",
    eprint = "1009.6100",
    archivePrefix = "arXiv",
    primaryClass = "nucl-th",
    doi = "10.1063/1.3541983",
    journal = "AIP Conf. Proc.",
    volume = "1322",
    number = "1",
    pages = "205--213",
    year = "2010"
}

@article{PavonValderrama:2019lsu,
    author = "Pav\'on Valderrama, Manuel",
    title = "{Scattering amplitudes versus potentials in nuclear effective field theory: is there a potential compromise?}",
    eprint = "1902.08172",
    archivePrefix = "arXiv",
    primaryClass = "nucl-th",
    month = "2",
    year = "2019"
}

@article{Mishra:2021luw,
    author = {Mishra, Chinmay and Ekstr\"om, A. and Hagen, G. and Papenbrock, T. and Platter, L.},
    title = "{Two-pion exchange as a leading-order contribution in chiral effective field theory}",
    eprint = "2111.15515",
    archivePrefix = "arXiv",
    primaryClass = "nucl-th",
    month = "11",
    year = "2021"
}

@article{Valderrama:2021bql,
    author = "Pav\'on Valderrama, Manuel",
    title = "{A comparison of two possible nuclear effective field theory expansions around the one- and two-pion exchange potentials}",
    eprint = "2112.02076",
    archivePrefix = "arXiv",
    primaryClass = "nucl-th",
    month = "12",
    year = "2021"
}

@article{Pandharipande:2005sx,
    author = "Pandharipande, V. R. and Phillips, Daniel R. and van Kolck, U.",
    title = "{Delta effects in pion-nucleon scattering and the strength of the two-pion-exchange three-nucleon interaction}",
    eprint = "nucl-th/0501061",
    archivePrefix = "arXiv",
    doi = "10.1103/PhysRevC.71.064002",
    journal = "Phys. Rev. C",
    volume = "71",
    pages = "064002",
    year = "2005"
}

@article{Pascalutsa:2002pi,
    author = "Pascalutsa, Vladimir and Phillips, Daniel R.",
    title = "{Effective theory of the delta(1232) in Compton scattering off the nucleon}",
    eprint = "nucl-th/0212024",
    archivePrefix = "arXiv",
    doi = "10.1103/PhysRevC.67.055202",
    journal = "Phys. Rev. C",
    volume = "67",
    pages = "055202",
    year = "2003"
}

@article{Long:2009wq,
    author = "Long, Bingwei and van Kolck, U.",
    title = "{pi N Scattering in the Delta(1232) Region in an Effective Field Theory}",
    eprint = "0907.4569",
    archivePrefix = "arXiv",
    primaryClass = "hep-ph",
    reportNumber = "ECT*-09-08, INT-PUB-09-036",
    doi = "10.1016/j.nuclphysa.2010.03.008",
    journal = "Nucl. Phys. A",
    volume = "840",
    pages = "39--75",
    year = "2010"
}

@article{Roper:1964zza,
    author = "Roper, L. David",
    title = "{Evidence for a P-11 Pion-Nucleon Resonance at 556 MeV}",
    doi = "10.1103/PhysRevLett.12.340",
    journal = "Phys. Rev. Lett.",
    volume = "12",
    pages = "340--342",
    year = "1964"
}

@article{Weinberg:1969hw,
    author = "Weinberg, Steven",
    title = "{Algebraic realizations of chiral symmetry}",
    doi = "10.1103/PhysRev.177.2604",
    journal = "Phys. Rev.",
    volume = "177",
    pages = "2604--2620",
    year = "1969"
}

@article{Beane:2002ud,
    author = "Beane, Silas R. and van Kolck, Ubirajara",
    title = "{The Role of the Roper in QCD}",
    eprint = "nucl-th/0212039",
    archivePrefix = "arXiv",
    reportNumber = "INT-PUB-02-52",
    doi = "10.1088/0954-3899/31/8/021",
    journal = "J. Phys. G",
    volume = "31",
    pages = "921--934",
    year = "2005"
}

@article{Borasoy:2006fk,
    author = "Borasoy, B. and Bruns, P. C. and Mei{\ss}ner, U.-G. and Lewis, R.",
    title = "{Chiral corrections to the Roper mass}",
    eprint = "hep-lat/0608001",
    archivePrefix = "arXiv",
    doi = "10.1016/j.physletb.2006.08.057",
    journal = "Phys. Lett. B",
    volume = "641",
    pages = "294--300",
    year = "2006"
}

@article{Djukanovic:2009gt,
    author = "Djukanovic, D. and Gegelia, J. and Scherer, S.",
    title = "{Chiral structure of the Roper resonance using complex-mass scheme}",
    eprint = "0903.0736",
    archivePrefix = "arXiv",
    primaryClass = "hep-ph",
    doi = "10.1016/j.physletb.2010.05.022",
    journal = "Phys. Lett. B",
    volume = "690",
    pages = "123--128",
    year = "2010"
}

@article{Peng:2020nyz,
    author = "Peng, Rui and Lyu, Songlin and Long, Bingwei",
    title = "{Perturbative chiral nucleon\textendash{}nucleon potential for the $^3P_0$ partial wave}",
    eprint = "2011.13186",
    archivePrefix = "arXiv",
    primaryClass = "nucl-th",
    doi = "10.1088/1572-9494/aba251",
    journal = "Commun. Theor. Phys.",
    volume = "72",
    number = "9",
    pages = "095301",
    year = "2020"
}

@article{Kaplan:2019znu,
    author = "Kaplan, David B.",
    title = "{Convergence of nuclear effective field theory with perturbative pions}",
    eprint = "1905.07485",
    archivePrefix = "arXiv",
    primaryClass = "nucl-th",
    reportNumber = "INT-PUB-19-015",
    doi = "10.1103/PhysRevC.102.034004",
    journal = "Phys. Rev. C",
    volume = "102",
    number = "3",
    pages = "034004",
    year = "2020"
}

@article{Fleming:1999ee,
    author = "Fleming, Sean and Mehen, Thomas and Stewart, Iain W.",
    title = "{NNLO corrections to nucleon-nucleon scattering and perturbative pions}",
    eprint = "nucl-th/9911001",
    archivePrefix = "arXiv",
    reportNumber = "UCSD-PTH-99-13, CALT-68-2243, UTPT-99-16",
    doi = "10.1016/S0375-9474(00)00221-9",
    journal = "Nucl. Phys. A",
    volume = "677",
    pages = "313--366",
    year = "2000"
}

@article{Wu:2018lai,
    author = "Wu, Shaowei and Long, Bingwei",
    title = "{Perturbative $NN$ scattering in chiral effective field theory}",
    eprint = "1807.04407",
    archivePrefix = "arXiv",
    primaryClass = "nucl-th",
    reportNumber = "CTP-SCU/2018002, CTP-SCU-2018002",
    doi = "10.1103/PhysRevC.99.024003",
    journal = "Phys. Rev. C",
    volume = "99",
    number = "2",
    pages = "024003",
    year = "2019"
}

@article{PavonValderrama:2016lqn,
    author = "Pav\'on Valderrama, M. and S\'anchez S\'anchez, M. and Yang, C. J. and Long, Bingwei and Carbonell, J. and van Kolck, U.",
    title = "{Power Counting in Peripheral Partial Waves: The Singlet Channels}",
    eprint = "1611.10175",
    archivePrefix = "arXiv",
    primaryClass = "nucl-th",
    doi = "10.1103/PhysRevC.95.054001",
    journal = "Phys. Rev. C",
    volume = "95",
    number = "5",
    pages = "054001",
    year = "2017"
}

@article{Birse:2005um,
    author = "Birse, Michael C.",
    title = "{Power counting with one-pion exchange}",
    eprint = "nucl-th/0507077",
    archivePrefix = "arXiv",
    doi = "10.1103/PhysRevC.74.014003",
    journal = "Phys. Rev. C",
    volume = "74",
    pages = "014003",
    year = "2006"
}

@article{Nogga:2005hy,
    author = "Nogga, A. and Timmermans, R. G. E. and van Kolck, U.",
    title = "{Renormalization of one-pion exchange and power counting}",
    eprint = "nucl-th/0506005",
    archivePrefix = "arXiv",
    reportNumber = "FZJ-IKP-TH-2005-19",
    doi = "10.1103/PhysRevC.72.054006",
    journal = "Phys. Rev. C",
    volume = "72",
    pages = "054006",
    year = "2005"
}

@article{Arndt:2006bf,
    author = "Arndt, R. A. and Briscoe, W. J. and Strakovsky, I. I. and Workman, R. L.",
    title = "{Extended partial-wave analysis of piN scattering data}",
    eprint = "nucl-th/0605082",
    archivePrefix = "arXiv",
    doi = "10.1103/PhysRevC.74.045205",
    journal = "Phys. Rev. C",
    volume = "74",
    pages = "045205",
    year = "2006"
}

@article{Bauer:2014cqa,
    author = "Bauer, T. and Scherer, S. and Tiator, L.",
    title = "{Electromagnetic transition form factors of the Roper resonance in effective field theory}",
    eprint = "1402.0741",
    archivePrefix = "arXiv",
    primaryClass = "nucl-th",
    reportNumber = "MITP-14-007",
    doi = "10.1103/PhysRevC.90.015201",
    journal = "Phys. Rev. C",
    volume = "90",
    number = "1",
    pages = "015201",
    year = "2014"
}

@article{Gegelia:2016xcw,
    author = "Gegelia, Jambul and Mei{\ss}ner, U.-G. and Yao, De-Liang",
    title = "{The width of the Roper resonance in baryon chiral perturbation theory}",
    eprint = "1606.04873",
    archivePrefix = "arXiv",
    primaryClass = "hep-ph",
    doi = "10.1016/j.physletb.2016.07.068",
    journal = "Phys. Lett. B",
    volume = "760",
    pages = "736--741",
    year = "2016"
}

@article{Gelenava:2017mmk,
    author = "Gelenava, Mishiko",
    title = "{Electromagnetic transition form factors of the Roper resonance in baryon chiral perturbation theory}",
    eprint = "1711.03494",
    archivePrefix = "arXiv",
    primaryClass = "nucl-th",
    doi = "10.1140/epja/i2018-12523-5",
    journal = "Eur. Phys. J. A",
    volume = "54",
    number = "5",
    pages = "88",
    year = "2018"
}

@article{Severt:2020jzc,
    author = "Severt, Daniel and Mei{\ss}ner, U.-G.",
    title = "{The Roper Resonance in a finite volume}",
    eprint = "2003.05745",
    archivePrefix = "arXiv",
    primaryClass = "hep-lat",
    doi = "10.1088/1572-9494/ab8a24",
    journal = "Commun. Theor. Phys.",
    volume = "72",
    number = "7",
    pages = "075201",
    year = "2020"
}

@article{Manohar:1983md,
    author = "Manohar, Aneesh and Georgi, Howard",
    title = "{Chiral Quarks and the Nonrelativistic Quark Model}",
    reportNumber = "HUTP-83/A042a",
    doi = "10.1016/0550-3213(84)90231-1",
    journal = "Nucl. Phys. B",
    volume = "234",
    pages = "189--212",
    year = "1984"
}

@article{vanKolck:2020plz,
    author = "van Kolck, U.",
    title = "{Naturalness in nuclear effective field theories}",
    eprint = "2003.09974",
    archivePrefix = "arXiv",
    primaryClass = "nucl-th",
    doi = "10.1140/epja/s10050-020-00092-1",
    journal = "Eur. Phys. J. A",
    volume = "56",
    number = "3",
    pages = "97",
    year = "2020"
}

@article{Machleidt:2011zz,
    author = "Machleidt, R. and Entem, D. R.",
    title = "{Chiral effective field theory and nuclear forces}",
    eprint = "1105.2919",
    archivePrefix = "arXiv",
    primaryClass = "nucl-th",
    doi = "10.1016/j.physrep.2011.02.001",
    journal = "Phys. Rept.",
    volume = "503",
    pages = "1--75",
    year = "2011"
}

@article{Epelbaum:2019kcf,
    author = "Epelbaum, Evgeny and Krebs, Hermann and Reinert, Patrick",
    title = "{High-precision nuclear forces from chiral EFT: State-of-the-art, challenges and outlook}",
    eprint = "1911.11875",
    archivePrefix = "arXiv",
    primaryClass = "nucl-th",
    doi = "10.3389/fphy.2020.00098",
    journal = "Front. in Phys.",
    volume = "8",
    pages = "98",
    year = "2020"
}

@article{Hammer:2019poc,
    author = {Hammer, H.-W. and K\"onig, S. and van Kolck, U.},
    title = "{Nuclear effective field theory: status and perspectives}",
    eprint = "1906.12122",
    archivePrefix = "arXiv",
    primaryClass = "nucl-th",
    doi = "10.1103/RevModPhys.92.025004",
    journal = "Rev. Mod. Phys.",
    volume = "92",
    number = "2",
    pages = "025004",
    year = "2020"
}

@article{Ren:2016jna,
    author = "Ren, Xiu-Lei and Li, Kai-Wen and Geng, Li-Sheng and Long, Bing-Wei and Ring, Peter and Meng, Jie",
    title = "{Leading order relativistic chiral nucleon-nucleon interaction}",
    eprint = "1611.08475",
    archivePrefix = "arXiv",
    primaryClass = "nucl-th",
    reportNumber = "CTP-SCU-2016012",
    doi = "10.1088/1674-1137/42/1/014103",
    journal = "Chin. Phys. C",
    volume = "42",
    number = "1",
    pages = "014103",
    year = "2018"
}

@article{Xiao:2018jot,
    author = "Xiao, Yang and Geng, Li-Sheng and Ren, Xiu-Lei",
    title = "{Covariant chiral nucleon-nucleon contact Lagrangian up to order $\mathcal{O}(q^4)$}",
    eprint = "1812.03005",
    archivePrefix = "arXiv",
    primaryClass = "nucl-th",
    doi = "10.1103/PhysRevC.99.024004",
    journal = "Phys. Rev. C",
    volume = "99",
    number = "2",
    pages = "024004",
    year = "2019"
}

@article{Xiao:2020ozd,
    author = "Xiao, Yang and Wang, Chun-Xuan and Lu, Jun-Xu and Geng, Li-Sheng",
    title = "{Two-pion exchange contributions to the nucleon-nucleon interaction in covariant baryon chiral perturbation theory}",
    eprint = "2007.13675",
    archivePrefix = "arXiv",
    primaryClass = "nucl-th",
    doi = "10.1103/PhysRevC.102.054001",
    journal = "Phys. Rev. C",
    volume = "102",
    number = "5",
    pages = "054001",
    year = "2020"
}

@article{Bai:2020yml,
    author = "Bai, Qian-Qian and Wang, Chun-Xuan and Xiao, Yang and Geng, Li-Sheng",
    title = "{Pion-mass dependence of the nucleon-nucleon interaction}",
    eprint = "2007.01638",
    archivePrefix = "arXiv",
    primaryClass = "nucl-th",
    doi = "10.1016/j.physletb.2020.135745",
    journal = "Phys. Lett.",
    volume = "B",
    pages = "135745",
    year = "2020"
}

@article{Wang:2020myr,
    author = "Wang, Chun-Xuan and Geng, Li-Sheng and Long, Bingwei",
    title = "{Renormalizability of leading order covariant chiral nucleon-nucleon interaction}",
    eprint = "2001.08483",
    archivePrefix = "arXiv",
    primaryClass = "nucl-th",
    doi = "10.1088/1674-1137/abe368",
    journal = "Chin. Phys. C",
    volume = "45",
    number = "5",
    pages = "054101",
    year = "2021"
}

@article{Ren:2017yvw,
    author = "Ren, Xiu-Lei and Wang, Chun-Xuan and Li, Kai-Wen and Geng, Li-Sheng and Meng, Jie",
    title = "{Relativistic Chiral Description of the 1 S 0 Nucleon\textendash{}Nucleon Scattering}",
    eprint = "1712.10083",
    archivePrefix = "arXiv",
    primaryClass = "nucl-th",
    doi = "10.1088/0256-307X/38/6/062101",
    journal = "Chin. Phys. Lett.",
    volume = "38",
    number = "6",
    pages = "062101",
    year = "2021"
}

@article{Bai:2021uim,
    author = "Bai, Qian-Qian and Wang, Chun-Xuan and Xiao, Yang and Lu, Jun-Xu and Geng, Li-Sheng",
    title = "{Nucleon-nucleon interaction in the S13-D13 coupled channel for a pion mass of 469 MeV}",
    eprint = "2105.06113",
    archivePrefix = "arXiv",
    primaryClass = "hep-ph",
    doi = "10.1016/j.physletb.2022.137347",
    journal = "Phys. Lett. B",
    volume = "833",
    pages = "137347",
    year = "2022"
}

@article{Lu:2021gsb,
    author = "Lu, Jun-Xu and Wang, Chun-Xuan and Xiao, Yang and Geng, Li-Sheng and Meng, Jie and Ring, Peter",
    title = "{Accurate Relativistic Chiral Nucleon-Nucleon Interaction up to Next-to-Next-to-Leading Order}",
    eprint = "2111.07766",
    archivePrefix = "arXiv",
    primaryClass = "nucl-th",
    doi = "10.1103/PhysRevLett.128.142002",
    journal = "Phys. Rev. Lett.",
    volume = "128",
    number = "14",
    pages = "142002",
    year = "2022"
}

@article{Wang:2021kos,
    author = "Wang, Chun-Xuan and Lu, Jun-Xu and Xiao, Yang and Geng, Li-Sheng",
    title = "{Nonperturbative two-pion exchange contributions to the nucleon-nucleon interaction in covariant baryon chiral perturbation theory}",
    eprint = "2110.05278",
    archivePrefix = "arXiv",
    primaryClass = "nucl-th",
    doi = "10.1103/PhysRevC.105.014003",
    journal = "Phys. Rev. C",
    volume = "105",
    number = "1",
    pages = "014003",
    year = "2022"
}

@article{Entem:2017gor,
    author = "Entem, D. R. and Machleidt, R. and Nosyk, Y.",
    title = "{High-quality two-nucleon potentials up to fifth order of the chiral expansion}",
    eprint = "1703.05454",
    archivePrefix = "arXiv",
    primaryClass = "nucl-th",
    doi = "10.1103/PhysRevC.96.024004",
    journal = "Phys. Rev. C",
    volume = "96",
    number = "2",
    pages = "024004",
    year = "2017"
}

@article{Kaiser:1997mw,
    author = "Kaiser, Norbert and Brockmann, R. and Weise, W.",
    title = "{Peripheral nucleon-nucleon phase shifts and chiral symmetry}",
    eprint = "nucl-th/9706045",
    archivePrefix = "arXiv",
    doi = "10.1016/S0375-9474(97)00586-1",
    journal = "Nucl. Phys. A",
    volume = "625",
    pages = "758--788",
    year = "1997"
}

@article{Kaiser:1998wa,
    author = "Kaiser, Norbert and Gerstend{\" o}rfer, S. and Weise, W.",
    title = "{Peripheral NN scattering: Role of delta excitation, correlated two pion and vector meson exchange}",
    eprint = "nucl-th/9802071",
    archivePrefix = "arXiv",
    doi = "10.1016/S0375-9474(98)00234-6",
    journal = "Nucl. Phys. A",
    volume = "637",
    pages = "395--420",
    year = "1998"
}

@article{Krebs:2007rh,
    author = "Krebs, Hermann and Epelbaum, Evgeny and Mei{\ss}ner, U.-G.",
    title = "{Nuclear forces with Delta-excitations up to next-to-next-to-leading order. I. Peripheral nucleon-nucleon waves}",
    eprint = "nucl-th/0703087",
    archivePrefix = "arXiv",
    reportNumber = "FZJ-IKP-TH-2007-9, HISKP-TH-07-06",
    doi = "10.1140/epja/i2007-10372-y",
    journal = "Eur. Phys. J. A",
    volume = "32",
    pages = "127--137",
    year = "2007"
}

@article{Epelbaum:2008td,
    author = "Epelbaum, E. and Krebs, H. and Mei{\ss}ner, U.-G.",
    title = "{Isospin-breaking two-nucleon force with explicit Delta-excitations}",
    eprint = "0801.1299",
    archivePrefix = "arXiv",
    primaryClass = "nucl-th",
    reportNumber = "FZJ-IKP-TH-2007-36, HISKP-TH-07-29",
    doi = "10.1103/PhysRevC.77.034006",
    journal = "Phys. Rev. C",
    volume = "77",
    pages = "034006",
    year = "2008"
}

@article{Piarulli:2014bda,
    author = "Piarulli, M. and Girlanda, L. and Schiavilla, R. and Navarro P\'erez, R. and Amaro, J. E. and Ruiz Arriola, E.",
    title = "{Minimally nonlocal nucleon-nucleon potentials with chiral two-pion exchange including $\Delta$ resonances}",
    eprint = "1412.6446",
    archivePrefix = "arXiv",
    primaryClass = "nucl-th",
    reportNumber = "JLAB-THY-14-1993",
    doi = "10.1103/PhysRevC.91.024003",
    journal = "Phys. Rev. C",
    volume = "91",
    number = "2",
    pages = "024003",
    year = "2015"
}

@article{Ekstrom:2017koy,
    author = {Ekstr\"om, A. and Hagen, G. and Morris, T. D. and Papenbrock, T. and Schwartz, P. D.},
    title = "{$\Delta$ isobars and nuclear saturation}",
    eprint = "1707.09028",
    archivePrefix = "arXiv",
    primaryClass = "nucl-th",
    doi = "10.1103/PhysRevC.97.024332",
    journal = "Phys. Rev. C",
    volume = "97",
    number = "2",
    pages = "024332",
    year = "2018"
}

@article{Strohmeier:2020dkb,
    author = "Strohmeier, Susanne and Kaiser, Norbert",
    title = "{Nucleon-Nucleon Scattering with Coupled Nucleon-Delta Channels in Chiral Effective Field Theory}",
    eprint = "2004.12964",
    archivePrefix = "arXiv",
    primaryClass = "nucl-th",
    doi = "10.1016/j.nuclphysa.2020.121980",
    journal = "Nucl. Phys. A",
    volume = "1002",
    pages = "121980",
    year = "2020"
}

@article{Nosyk:2021pxb,
    author = "Nosyk, Y. and Entem, D. R. and Machleidt, R.",
    title = "{Nucleon-nucleon potentials from \ensuremath{\Delta}-full chiral effective-field-theory and implications}",
    eprint = "2107.06452",
    archivePrefix = "arXiv",
    primaryClass = "nucl-th",
    doi = "10.1103/PhysRevC.104.054001",
    journal = "Phys. Rev. C",
    volume = "104",
    number = "5",
    pages = "054001",
    year = "2021"
}

@article{Epelbaum:2007sq,
    author = "Epelbaum, E. and Krebs, H. and Mei{\ss}ner, U.-G.",
    title = "{Delta-excitations and the three-nucleon force}",
    eprint = "0712.1969",
    archivePrefix = "arXiv",
    primaryClass = "nucl-th",
    reportNumber = "FZJ-IKP-TH-2007-34, HISKP-TH-07-28",
    doi = "10.1016/j.nuclphysa.2008.02.305",
    journal = "Nucl. Phys. A",
    volume = "806",
    pages = "65--78",
    year = "2008"
}

@article{Kaiser:2015yca,
    author = "Kaiser, N.",
    title = "{Three-pion exchange nucleon-nucleon potentials with virtual \ensuremath{\Delta}-isobar excitation}",
    eprint = "1504.05131",
    archivePrefix = "arXiv",
    primaryClass = "nucl-th",
    doi = "10.1103/PhysRevC.92.024002",
    journal = "Phys. Rev. C",
    volume = "92",
    number = "2",
    pages = "024002",
    year = "2015"
}

@article{Krebs:2018jkc,
    author = "Krebs, H. and Gasparyan, A. M. and Epelbaum, E.",
    title = "{Three-nucleon force in chiral EFT with explicit $\Delta(1232)$ degrees of freedom: Longest-range contributions at fourth order}",
    eprint = "1803.09613",
    archivePrefix = "arXiv",
    primaryClass = "nucl-th",
    doi = "10.1103/PhysRevC.98.014003",
    journal = "Phys. Rev. C",
    volume = "98",
    number = "1",
    pages = "014003",
    year = "2018"
}

@article{Logoteta:2016nzc,
    author = "Logoteta, Domenico and Bombaci, Ignazio and Kievsky, Alejandro",
    title = "{Nuclear matter properties from local chiral interactions with $\Delta$ isobar intermediate states}",
    eprint = "1609.00649",
    archivePrefix = "arXiv",
    primaryClass = "nucl-th",
    doi = "10.1103/PhysRevC.94.064001",
    journal = "Phys. Rev. C",
    volume = "94",
    number = "6",
    pages = "064001",
    year = "2016"
}

@article{Long:2011rt,
    author = "Long, Bingwei and van Kolck, U.",
    title = "{The Role of the Roper in Chiral Perturbation Theory}",
    eprint = "1105.2764",
    archivePrefix = "arXiv",
    primaryClass = "nucl-th",
    reportNumber = "JLAB-THY-11-1372, INT-PUB-11-020",
    doi = "10.1016/j.nuclphysa.2011.09.002",
    journal = "Nucl. Phys. A",
    volume = "870-871",
    pages = "72--82",
    year = "2011"
}

@article{Stapp:1956mz,
    author = "Stapp, H. P. and Ypsilantis, T. J. and Metropolis, N.",
    title = "{Phase shift analysis of 310-MeV proton proton scattering experiments}",
    doi = "10.1103/PhysRev.105.302",
    journal = "Phys. Rev.",
    volume = "105",
    pages = "302--310",
    year = "1957"
}

@article{Jenkins:1990jv,
    author = "Jenkins, Elizabeth Ellen and Manohar, Aneesh V.",
    title = "{Baryon chiral perturbation theory using a heavy fermion Lagrangian}",
    reportNumber = "UCSD-PTH-90-23",
    doi = "10.1016/0370-2693(91)90266-S",
    journal = "Phys. Lett. B",
    volume = "255",
    pages = "558--562",
    year = "1991"
}

@article{Jenkins:1991es,
    author = "Jenkins, Elizabeth Ellen and Manohar, Aneesh V.",
    title = "{Chiral corrections to the baryon axial currents}",
    reportNumber = "UCSD-PTH-91-05",
    doi = "10.1016/0370-2693(91)90840-M",
    journal = "Phys. Lett. B",
    volume = "259",
    pages = "353--358",
    year = "1991"
}

@article{Lomon:1981su,
    author = "Lomon, Earle L.",
    title = "{Coupled Channel Interaction for Intermediate-energy Nucleon-nucleon Reactions}",
    reportNumber = "MIT-CTP-963",
    doi = "10.1103/PhysRevD.26.576",
    journal = "Phys. Rev. D",
    volume = "26",
    pages = "576",
    year = "1982"
}

@article{Lee:1984us,
    author = "Lee, T.-S. H.",
    title = "{MESON THEORY OF NUCLEON NUCLEON SCATTERING UP TO 2-GEV}",
    doi = "10.1103/PhysRevC.29.195",
    journal = "Phys. Rev. C",
    volume = "29",
    pages = "195--203",
    year = "1984"
}

@article{Ray:1987ir,
    author = "Ray, L.",
    title = "{Nucleon Nucleon Scattering With Isobar Degrees of Freedom}",
    doi = "10.1103/PhysRevC.35.1072",
    journal = "Phys. Rev. C",
    volume = "35",
    pages = "1072--1082",
    year = "1987"
}

@article{Sitarski:1987gd,
    author = "Sitarski, W. P. and Blunden, P. G. and Lomon, E. L.",
    title = "{Deuteron Properties of the Coupled Nucleon and Isobar Channels Model}",
    reportNumber = "MIT-CTP-1407",
    doi = "10.1103/PhysRevC.36.2479",
    journal = "Phys. Rev. C",
    volume = "36",
    pages = "2479--2494",
    year = "1987"
}

@article{Bernard:1996gq,
    author = "Bernard, V. and Kaiser, Norbert and Mei{\ss}ner, Ulf-G.",
    title = "{Aspects of chiral pion - nucleon physics}",
    eprint = "hep-ph/9611253",
    archivePrefix = "arXiv",
    reportNumber = "KFA-IKP-TH-1996-14, KFA-IKP-TH-1996-22, LPT-96-22",
    doi = "10.1016/S0375-9474(97)00021-3",
    journal = "Nucl. Phys. A",
    volume = "615",
    pages = "483--500",
    year = "1997"
}

@article{Friar:1999sj,
    author = "Friar, James Lewis",
    title = "{Equivalence of nonstatic two pion exchange nucleon-nucleon potentials}",
    eprint = "nucl-th/9901082",
    archivePrefix = "arXiv",
    reportNumber = "LA-UR-99-296, LA-UR-99-296-REV",
    doi = "10.1103/PhysRevC.60.034002",
    journal = "Phys. Rev. C",
    volume = "60",
    pages = "034002",
    year = "1999"
}

@article{Stoks:1993tb,
    author = "Stoks, V. G. J. and Klomp, R. A. M. and Rentmeester, M. C. M. and de Swart, J. J.",
    title = "{Partial wave analaysis of all nucleon-nucleon scattering data below 350-MeV}",
    doi = "10.1103/PhysRevC.48.792",
    journal = "Phys. Rev. C",
    volume = "48",
    pages = "792--815",
    year = "1993"
}

@article{Kaiser:2026lvl,
    author = "Kaiser, Norbert",
    title = "{Two-pion exchange nucleon-nucleon potentials with Roper resonance excitation}",
    eprint = "2602.18664",
    archivePrefix = "arXiv",
    primaryClass = "nucl-th",
    month = "2",
    year = "2026"
}

@article{Wang:2018atz,
    author = "Wang, Bo and Liu, Zhan-Wei and Liu, Xiang",
    title = "{$\bar{B}^{(\ast)} \bar{B}^{(\ast)}$ interactions in chiral effective field theory}",
    eprint = "1812.04457",
    archivePrefix = "arXiv",
    primaryClass = "hep-ph",
    doi = "10.1103/PhysRevD.99.036007",
    journal = "Phys. Rev. D",
    volume = "99",
    number = "3",
    pages = "036007",
    year = "2019"
}

@article{Wang:2019ato,
    author = "Wang, Bo and Meng, Lu and Zhu, Shi-Lin",
    title = "{Hidden-charm and hidden-bottom molecular pentaquarks in chiral effective field theory}",
    eprint = "1909.13054",
    archivePrefix = "arXiv",
    primaryClass = "hep-ph",
    doi = "10.1007/JHEP11(2019)108",
    journal = "JHEP",
    volume = "11",
    pages = "108",
    year = "2019"
}

\end{document}